\documentclass[aps,prd,twocolumn,groupedaddress,floatfix,superscriptaddress]{revtex4-1}
\usepackage{lipsum}
\usepackage{capt-of}
\usepackage{bm}% bold math
\usepackage{amssymb}
\usepackage{amsmath}
\usepackage{graphicx}
\usepackage{epsfig}
\usepackage{array}
\usepackage{float}
\usepackage{color}
\usepackage[usenames,dvipsnames]{xcolor}
\usepackage{epstopdf}
\usepackage{natbib}
\usepackage{lineno}
\usepackage{slashed}
\usepackage{color}
\usepackage{caption}
\usepackage{hyperref}
\hypersetup{colorlinks=true, linkcolor=NavyBlue, citecolor=blue,urlcolor=Blue}
\usepackage{slashed}
\usepackage{soul}
\usepackage[bottom]{footmisc}
\usepackage{wrapfig}
\usepackage{mathtools,slashed}

\newcommand{\bea}{\begin{eqnarray}}
\newcommand{\eea}{\end{eqnarray}}
\newcommand{\be}{\begin{equation}}
\newcommand{\ee}{\end{equation}}

\newcommand{\sgn}{{\rm sign}}

\renewcommand\Re{\text{Re}}
\renewcommand\Im{\text{Im}}
\newcommand{\nn}{\nonumber}
\newcommand{\y}{y_\text{T}}
\begin{document}

\title{Fermion mass and width in QED in a magnetic field}
%\affiliation{Facultad de F\'isica, Pontificia Universidad Cat\'olica de Chile, Vicu\~{n}a Mackenna 4860, Santiago, Chile}
\author{Alejandro Ayala}
\affiliation{Instituto de Ciencias
  Nucleares, Universidad Nacional Aut\'onoma de M\'exico, Apartado
  Postal 70-543, CdMx 04510,
  Mexico.}
 \affiliation{Centre for Theoretical and Mathematical Physics, and Department of Physics, University of Cape Town, Rondebosch 7700, South Africa.}
\author{Jorge David Casta\~no-Yepes}
  \affiliation{{Facultad de Ciencias - CUICBAS, Universidad de Colima, Bernal D\'iaz del Castillo No. 340, Col. Villas San Sebasti\'an, 28045 Colima, Mexico.}}
\author{M. Loewe}
\affiliation{Instituto de F\'{\i}sica, Pontificia Universidad Cat\'olica de Chile, Casilla 306, Santiago 22, Chile.}
 \affiliation{Centre for Theoretical and Mathematical Physics, and Department of Physics, University of Cape Town, Rondebosch 7700, South Africa.}
\affiliation{Centro Cient\'\i fico-Tecnol\'ogico de Valpara\'\i so CCTVAL, Universidad T\'ecnica Federico Santa Mar\'\i a, Casilla 110-V, Valapara\'\i so, Chile}
\author{Enrique Mu\~noz}
\affiliation{Instituto de F\'{\i}sica, Pontificia Universidad Cat\'olica de Chile, Casilla 306, Santiago 22, Chile.}
\affiliation{Research Center for Nanotechnology and Advanced Materials CIEN-UC, Pontificia Universidad Católica de Chile, Santiago, Chile.}

% \begin{abstract}
% Some mathematical details about the steepest-descent calculations in the previous notes
% \end{abstract}
\begin{abstract}

We revisit the calculation of the fermion self-energy in QED in the presence of a magnetic field. We show that, after carrying out the renormalization procedure and identifying the most general {perturbative} tensor structure for the modified fermion {mass operator} in the large field limit, the mass develops an imaginary part. This happens when account is made of the sub-leading contributions associated to Landau levels other than the lowest one. The imaginary part is associated to a spectral density describing the spread of the mass function in momentum. The center of the distribution corresponds to the magnetic-field modified mass. The width becomes small as the field intensity increases in such a way that for asymptotically large values of the field, when the separation between Landau levels becomes also large, the mass function describes a stable particle occupying only the lowest Landau level. For large but finite values of the magnetic field, the spectral density represents a finite probability for the fermion to occupy Landau levels other than the LLL.

\end{abstract}
\pacs{uu}
%\pacs{...}
\maketitle
\maketitle

\section{Introduction}\label{SecI}

Magnetic fields influence the propagation properties of electrically charged as well as of neutral particles. Whereas charged particles couple directly to the magnetic field, neutral particles are affected indirectly when their quantum fluctuations involve charged particles. For instance, in QED, the coupling of the magnetic field to photon charged fluctuations gives rise to vacuum birefringence, whereby photons develop { three polarization modes for the two polarization states. This means that the breaking of Lorentz invariance due to the external magnetic field, implies the appareance of the three tensor structures that span the polarization tensor. As a consequence, the refractive index depends on the coefficient (mode) of each projection as well as on the propagation direction}~\cite{Hattori:2012je,PhysRevD.101.036016,Ayala:2020wzl}. 

In the same manner, fermions are also influenced by magnetic fields through quantum fluctuations that involve charged particles. This influence is encoded in the fermion self-energy. Attention to this object has been payed since the pioneering work by Schwinger~\cite{Schwinger:1951nm} who computed the fermion propagator in the presence of a uniform external magnetic field. In particular,
the use of non-perturbative techniques revealed the magnetic catalysis phenomenon, whereby a magnetic field of arbitrary intensity is able to dynamically generate a fermion mass, even when starting from massless fermions~\cite{Gusynin:1995nb,Leung:1995mh,Gusynin:1995gt}.  

In vacuum, perturbative calculations have focused on finding the leading magnetic corrections to the fermion mass for strong fields~\cite{Tsai:1974df,Jancovici:1970ep,Dittrich:1985yb,Machet:2015swa,Gepraegs:1994hy}. After renormalization, these corrections turn out proportional to $\left[\ln|eB|/m^2\right]^2$, where $|eB|$ and $m$ are the strength of the magnetic field and the fermion mass, respectively. These {\it double logarithmic} corrections become large when the ratio $|eB|/m^2$ is large, which happens for either large field intensities or small fermion masses, signaling the need of resummation. Carrying out this program, Ref.~\cite{Gusynin:1998nh} studied the transition between the perturbative and non-perturbative domains. A slightly different approach, where the effects of the magnetic field-induced photon polarization are included, is studied in Ref.~\cite{Kuznetsov:2002qb}. 

In all these works, the calculations focused on finding the kinematical domain of integration that leads to the double logarithms for strong fields, which essentially corresponds to finding the contribution from the lowest Landau level (LLL) { for the internal fermion line}. However, it is well-known that for large but still finite field strengths, the contribution of levels other than the LLL also become important. Furthermore, in order to find the magnetic field-driven mass corrections, {several} calculations { \cite{Tsai:1974df,Jancovici:1970ep,Dittrich:1985yb,Machet:2015swa,Gepraegs:1994hy}} resort to finding the self-energy matrix element in the fermion {\it preferred state}, namely, the spinor representing the lowest energy fermion state. For these purposes, Schwinger's phase factor is kept all along the calculation, which makes it a bit more cumbersome since then, the kinematical momentum $\Pi^\mu = p^\mu - eA^\mu (x)$, instead of the canonical momentum $p^\mu$, is involved, where $A^\mu(x)$ is the four-vector potential that gives rise to the magnetic field. Nevertheless, as is also well-known, for the self-energy calculation, the phase factor can be {\it gauged away} by choosing an appropriate gauge transformation. One can then ask whether the magnetic field induced fermion mass can be found from an approach where this phase is gauged away and  that emphasizes the general tensor structure of the fermion self-energy, from whose coefficients the magnetic field dependent fermion mass can be read off. 

In this work we revisit the calculation of the fermion self-energy in the presence of an external magnetic field. We carry out the  computation of the magnetic field induced fermion self-energy and then, from its general structure, we find the mass and width. After introducing a Schwinger parametrization and removing the vacuum, we identify the three distinct regions of integration that contribute to the mass shift. We show that, while working in the large field limit, it is still possible to include the effect of all Landau levels. The procedure we employ makes it also possible to find the subdominant contributions in the ratio $|eB|/m^2$ which include, in particular, the imaginary part of the self-energy. We interpret the result in terms of the development not only of a  mass but also of a width. For large field strengths, the former comes mainly from the Lowest Landau level (LLL) whereas the latter comes mainly from contributions other than the LLL. The work is organized as follows: In Sec.~\ref{SecII} we set up the computation of the one-loop fermion self-energy in the presence of a constant magnetic field. In Sec.~\ref{SecIII} we implement the requirements of renormalization in vacuum and find the general expression for the renormalized, magnetic field dependent self-energy. In Sec.~\ref{SecIV} we express the  mass shift operators in terms of its most general tensor structure. In order to find the explicit result, we separate the integration domain into the three distinct regions that contribute. {We show that one of these regions provides the dominant, real contribution, as well as a subleading imaginary part, implying a finite decay rate, whereas the other two regions contribute with subdominant terms}. In Sec.~\ref{SecV} we analyze the results expressing the self-energy in terms of its most general tensor structure and for different values of the magnetic field strength comparing to those obtained from considering only the leading, double logarithmic contribution. We show that the result can be understood as a mass shift with a Lorentzian width. As the magnetic field increases, the width decreases in such a way that for very large magnetic fields the Lorentzian turns into a Dirac delta function expressing the mass shift and dominated by the LLL. We finally summarize and conclude in Sec.~\ref{Concl} and leave for the appendices the explicit computation of some of the expressions and calculations that we use throughout the rest of the work.

\section{Self-energy}\label{SecII}

We start by considering the expression for the fermion self-energy in QED, at one loop, in the presence of an external magnetic field directed alog the $\hat{z}$-axis, namely, $\mathbf{B} = \hat{z} B$,  
\begin{eqnarray}
\!\!\!-i \Sigma(p) = \left(-i e \right)^2 \int\frac{d^4 k}{(2\pi)^4}\gamma^{\mu} i S_{F}(k) \gamma^{\nu} G_{\nu\mu}( p - k).
\label{selfenergy}
\end{eqnarray}

Here, we use the photon propagator in the Feynman gauge
\begin{eqnarray}
G_{\nu\mu}(p-k)&=&\frac{-i g_{\mu\nu}}{(p - k)^2 + i\epsilon}\nn\\
&=& - g_{\mu\nu}\int_{0}^{\infty}dx\,e^{i x\left[(p - k)^2 + i\epsilon\right]}, 
\label{eq:SP}
\end{eqnarray}
and the fermion propagator in the presence of the magnetic field, using Schwinger's proper-time representation
\begin{eqnarray}
S_{F}(k) &=& -i \int_{0}^{\infty} \frac{d\tau}{\cos(e B \tau)}e^{i \tau \left(k_{\parallel}^2 -
k_{\perp}^2\frac{\tan(e B \tau)}{e B \tau} - m^2 + i\epsilon \right)}\nn\\
&\times&
\left\{\left[\cos(e B \tau) + i \gamma^1 \gamma^2 \sin(e B \tau) \right]\left( m + \slashed{k}_{\parallel} \right)\right.\nn\\
&+&\left. \frac{\slashed{k}_{\perp}}{\cos(e B \tau)}
\right\},
\label{Eq:Invariant_propagator}
\end{eqnarray}
where the phase factor has been gauged away (see appendix~\ref{AppA} for details).

In order to separate the two directions, {\it i.e.} parallel and perpendicular, with respect to the magnetic field, we adopt the following conventional definitions: For the metric tensor $g^{\mu\nu} = g^{\mu\nu}_{\parallel} + g^{\mu\nu}_{\perp}$, with
\begin{subequations}
\bea
g_{\parallel}^{\mu\nu}=\text{diag}(1,0,0,-1),
\eea
and
\bea
g_{\perp}^{\mu\nu}=\text{diag}(0,-1,-1,0).
\eea
\end{subequations}

Therefore, we have the basic relations
\begin{subequations}
\begin{eqnarray}
\slashed{k} = \slashed{k}_{\parallel} + \slashed{k}_{\perp}
\end{eqnarray}
and
\begin{eqnarray}
k^2= k_{\parallel}^2 - k_{\perp}^2,
\end{eqnarray}
\end{subequations}
with $k_{\parallel}^2 = k_0^2 - k_3^2$ and $k_{\perp}^2 = k_1^2 + k_2^2$.

By applying the elementary properties of the algebra of Dirac matrices, we obtain (see details in  Appendix~\ref{AppB})
\begin{eqnarray}
\gamma^{\mu} i S_{F}(k)\gamma_{\mu} &=& \int_{0}^{\infty} \frac{d\tau}{\cos(e B \tau)}e^{i \tau \left(k_{\parallel}^2 -
k_{\perp}^2\frac{\tan(e B \tau)}{e B \tau} - m^2 + i\epsilon \right)}\nonumber\\
&\times&
\Big\{
4 m \cos(e B \tau) - 2\slashed{k}_{\parallel}\cos(e B \tau)\nonumber\\
&-&i \sin(e B \tau) \gamma^{1}\gamma^{2}\slashed{k}_{\parallel}
- \frac{2\slashed{k}_{\perp}}{\cos(e B \tau)}
\Big\}.
\label{eq:gamSFgam}
\end{eqnarray}

Inserting Eq.~(\ref{eq:SP}) and Eq.~(\ref{eq:gamSFgam}) into Eq.~(\ref{selfenergy}), the global exponential factor becomes
\begin{widetext}
\begin{eqnarray}
e^{\left[i\left\{x\left(\left(p - k \right)^2 + i\epsilon  \right) + \tau\left( k_{\parallel}^2 -
k_{\perp}^2\frac{\tan(e B \tau)}{eB \tau} - m^2 + i\epsilon\right)  \right\}\right]}=e^{\left[ i\left\{ 
(x + \tau) l_{\parallel}^2 - \left(x + \frac{\tan(e B \tau)}{e B} \right)l_{\perp}^2 + x p^2
- \frac{x^2}{x + \tau} p_{\parallel}^2 + \frac{x^2}{x + \frac{\tan(e B \tau)}{e B}} p_{\perp}^2 - m^2\tau + i\epsilon
\right\}\right]},
\end{eqnarray}
\end{widetext}
where we defined the shifted, internal momentum variables
\begin{eqnarray}
l_{\parallel}^{\mu} &=&  k_{\parallel}^{\mu} - \frac{x}{x + \tau}p_{\parallel}^{\mu}\\
l_{\perp}^{\mu} &=&  k_{\perp}^{\mu} - \frac{x}{x + \frac{\tan(e B\tau)}{e B}}p_{\perp}^{\mu}
\end{eqnarray}

After integrating over the internal momenta, using the simple identities
\begin{eqnarray}
\!\!\!\!\!\!\!\!\!\!\int \frac{d^2 l_{\perp}}{(2\pi)^2}
e^{-i\left(x + \frac{\tan(e B \tau)}{e B} \right)l_{\perp}^2} &=& \frac{1}{(2\pi)^2}\frac{- i \pi}{x + \frac{\tan(e B \tau)}{e B}},\\
\!\!\!\!\!\!\!\!\!\!\int \frac{d^2 l_{\perp}}{(2\pi)^2} l_{\perp}^{\mu}
e^{-i\left(x + \frac{\tan(e B \tau)}{e B} \right)l_{\perp}^2} &=& 0,\\
\!\!\!\!\!\!\!\!\!\!\int\frac{d^2 l_{\parallel}}{(2\pi)^2} e^{i(x + \tau)l_{\parallel}^2} &=& \frac{1}{(2\pi)^2}\frac{\pi}{(x + \tau)},\\
\!\!\!\!\!\!\!\!\!\!\int\frac{d^2 l_{\parallel}}{(2\pi)^2}l_{\parallel}^{\mu} e^{i(x + \tau)l_{\parallel}^2} &=& 0,
\end{eqnarray}
we obtain the expression
\begin{eqnarray}
\Sigma(p,B) &=& \frac{2 e^2}{(4 \pi)^2}\int_{0}^{\infty} \int_{0}^{\infty}\frac{dxd\tau}{(x + \tau)\left(x + \frac{\tan(e B \tau)}{e B} \right)}\nonumber\\
&\times&
\left[ 2m - \frac{x}{x + \tau}\slashed{p}_{\parallel} - \frac{x \slashed{p}_{\perp}}{\left(x + \frac{\tan(e B \tau)}{e B} \right)\left[\cos(e B \tau)\right]^2}\right.\nonumber\\
&-&\left. \frac{x\tan(e B \tau)}{x + \tau}i \gamma^1 \gamma^2 \slashed{p}_{\parallel} \right]\nonumber\\
&\times&e^{i\left(x p^2 - \frac{x^2}{x + \tau} p_{\parallel}^2 + \frac{x^2 p_{\perp}^{2}}{x + \frac{\tan(e B \tau)}{e B}} - \tau m^2 + i\epsilon\right)}.
\end{eqnarray}

Introducing the change of variables
\begin{eqnarray}
\tau &=& \frac{s(1 - y)}{m^2},\nonumber\\
x &=& \frac{s y }{m^2},\nonumber\\
\mathcal{B} &=& \frac{|e B|}{m^2},\nonumber\\
\rho_{\perp,\parallel}^2 &=& \frac{p_{\perp,\parallel}^2}{m^2},
\end{eqnarray}
with the corresponding Jacobian
\begin{eqnarray}
\frac{\partial(\tau,x)}{\partial(s,y)} = \left|\begin{array}{cc}1 - y & -s\\y & s \end{array}\right| = s,
\end{eqnarray}
we obtain that the self-energy is expressed by
\begin{eqnarray}
\Sigma(p,B)&=& \frac{2m e^2}{(4\pi)^2}\int_{0}^{\infty}\frac{ds}{s}\int_{0}^{1}dy\left[
(A) + (B) - (C)
\right]\nonumber\\ 
&\times&
e^{i s \left( \varphi(y,\rho,B)  + i\epsilon\right)},
\label{eq:selfBunr}
\end{eqnarray}
where we defined the phase
\begin{eqnarray}
\varphi(y,\rho,B)&=&y \rho^2 - y^2 \rho_{\parallel}^2\nonumber\\
&+& \frac{y^2 \cos(\mathcal{B}s(1-y))\rho_{\perp}^2}{y\cos(\mathcal{B}s(1-y)) + \frac{\sin(\mathcal{B}s(1-y))}{\mathcal{B}s}}\nonumber\\ 
&-& (1 - y),
\label{eq:phaseB}
\end{eqnarray}
as well as the terms that appear in the integrand
\begin{eqnarray}
(A) &=& \frac{(2 - y\slashed{\rho}_{\parallel})\cos(\mathcal{B}s(1-y))}{y \cos(\mathcal{B}s(1-y)) + \frac{\sin(\mathcal{B}s(1-y))}{\mathcal{B}s}},
\label{TermA}
\eea
\bea
(B) &=& \frac{-y \slashed{\rho}_{\perp}}{\left[y\cos(\mathcal{B}s(1-y)) + \frac{\sin(\mathcal{B}s(1-y))}{\mathcal{B}s} \right]^2},
\label{termB}
\eea
\bea
(C) &=& \frac{y\sin(\mathcal{B}s(1-y))}{{y\cos(\mathcal{B}s(1-y)) + \frac{\sin(\mathcal{B}s(1-y))}{\mathcal{B}s}} }\nonumber\\
&\times&i \gamma^1 \gamma^2\sgn(eB) \slashed{\rho}_{\parallel}.
\label{TermC}
\end{eqnarray}

\section{Fixing the counterterms in the $B=0$ limit}\label{SecIII}
Equation~(\ref{eq:selfBunr}) corresponds to the unrenormalized self-energy
for arbitrary magnetic field intensities. We impose the renormalization conditions such that $m$ corresponds to the physical mass at $B = 0$, {\it i.e.}
\begin{eqnarray}
\left.\Sigma^{ren}(p,0)\right|_{\slashed{p} = m} = 0,
\label{eq:R1}
\end{eqnarray}
and the corresponding condition for the wavefunction renormalization factor
\begin{eqnarray}
\left.\frac{\partial}{\partial\slashed{p}} \Sigma^{ren}(p,0)\right|_{\slashed{p} = m} = 0.
\label{eq:R2}
\end{eqnarray}

Each of these conditions will determine a counter-term to be added to the integrand in Eq.~ (\ref{eq:selfBunr}). { Note that the expressions for the counter-terms are given by imposing conditions over the canonical momentum $\slashed{p}$ instead of the kinematical one $\slashed{\Pi}$. Nevertheless, both prescriptions are identical, given that $\slashed{\Pi}\to\slashed{p}$ when $B\to0$, and hence, for the purpose of fixing the appropriate counter-terms, the renormalization conditions can be equivalently expressed in terms of the canonical momentum. Of course, the coincidence between prescriptions comes form the fact that the potential $A^\mu$ in the minimal coupling ($\Pi^\mu=p^\mu-eA^\mu$) vanishes at zero magnetic field in Mikowski space. However, this is not the more general case: in curved spaces, the appearance of a pseudo-magnetic field is possible, which arises from curvature effects (see for example Ref.~\cite{PhysRevB.95.125432}).}
%, but although it seems interesting, the deviations from flat-space are out of the scope of the present manuscript.}

{ Coming back to the counter-terms calculation, let us start with the phase defined in Eq.~(\ref{eq:phaseB}), by obtaining the limit at $B = 0$, namely,}
\begin{eqnarray}
\lim_{B\rightarrow0} \varphi(y,\rho,B) \equiv \varphi(y,\rho,0) &=& y\rho^2 - y^2\left(\rho_{\parallel}^2 - \rho_{\perp}^2 \right)\nonumber\\ 
&-& (1-y)\nonumber\\
&=& (1 - y)(y\rho^2 - 1)\nonumber\\
&=& (1 - y)y(\slashed{\rho}^2 - 1)\nonumber\\
&-&(1 - y)^2.
\label{eq:phase0}
\end{eqnarray}
For the terms in the integrand
\begin{eqnarray}
\lim_{B\rightarrow 0}(A)&=& 2 - y\slashed{\rho}_{\parallel}\nonumber\\
\lim_{B\rightarrow 0}(B)&=& -y\slashed{\rho}_{\perp}\nonumber\\
\lim_{B\rightarrow 0}(C)&=& 0.
\label{eq:ABC0}
\end{eqnarray}

Therefore, using $\slashed{\rho} = \slashed{\rho}_{\parallel} + \slashed{\rho}_{\perp}$, we have for the self-energy at $B=0$
\begin{eqnarray}
\Sigma^{ren}(p,0)&=& \frac{2m e^2}{(4\pi)^2}\int_{0}^{\infty}\frac{ds}{s}\int_{0}^{1}dye^{i s \left( -(1-y)^2  + i\epsilon\right)}\nonumber\\
&\times&
\left[
\left(2 - y\slashed{\rho}\right)e^{i sy(1-y)(\slashed{\rho}^2 -1)} + c.t.
\right],
\end{eqnarray}
where \lq\lq {\it c.t.}" represents the counterterms to be added to satisfy Eq.~(\ref{eq:R1})
and Eq.~(\ref{eq:R2}), respectively. In terms of the dimensionless variable
$\slashed{\rho} = \slashed{p}/m$, the condition of Eq.~(\ref{eq:R1}) becomes
\begin{eqnarray}
\Sigma^{ren}(p,0)|_{\slashed{p} = m} = \Sigma^{ren}(p,0)|_{\slashed{\rho} = 1} = 0,
\end{eqnarray}
and hence we conclude that the corresponding counterterm is
\begin{eqnarray}
c.t._1 =  - (2 - y)
\label{eq:ct1}
\end{eqnarray}

On the other hand, for the condition Eq.~(\ref{eq:R2}) we consider the first derivative of the (unrenormalized) self-energy on shell
\begin{eqnarray}
\left.\frac{\partial}{\partial\slashed{p}}\Sigma(p,m)\right|_{\slashed{p}=m}
&=& \frac{1}{m}\left.\frac{\partial}{\partial\slashed{\rho}}\Sigma(p,m)\right|_{\slashed{\rho}=1}\nonumber\\
&=& \frac{2m e^2}{(4\pi)^2}\!\int_{0}^{\infty}\!\frac{ds}{s}\int_{0}^{1}\!dye^{i s \left( -(1-y)^2+ i\epsilon\right)}\nonumber\\
&\times&\left[
-\frac{y}{m} e^{i sy(1-y)(\slashed{\rho}^2 -1)}\right.\nonumber\\
&+&\left. 
\left(2 - y\slashed{\rho}\right)
2 i s y(1-y)\right.\nonumber\\
&\times&\left.\slashed{\rho}e^{i sy(1-y)(\slashed{\rho}^2 -1)}\right]_{\slashed{\rho}=1}\nonumber\\
&=& \frac{2m e^2}{(4\pi)^2}\int_{0}^{\infty}\frac{ds}{s}\int_{0}^{1}dye^{i s \left( -(1-y)^2+ i\epsilon\right)}\nonumber\\
&\times&
\left[
-\frac{y}{m}  + 2 i s \frac{y(1-y)}{m} 
\left(2 - y\right)\right].
\end{eqnarray}

Thus, the second counterterm is
\begin{eqnarray}
c.t._2 = -\left(\slashed{\rho} - 1 \right)\left\{
-\frac{y}{m}  + 2 i s \frac{y(1-y)}{m} 
\left(2 - y\right)
\right\}.
\label{eq:ct2}
\end{eqnarray}

In summary, including both counterterms the renormalized self-energy at $B=0$ is given by the expression
\begin{eqnarray}
\Sigma^{ren}(p,0) &=& \frac{2m e^2}{(4\pi)^2}\int_{0}^{\infty}\frac{ds}{s}\int_{0}^{1}dye^{i s \left( -(1-y)^2  + i\epsilon\right)}\nonumber\\
&\times&
\left[
\left(2 - y\slashed{\rho}\right)e^{i s \left(\varphi(y,\rho,0) + (1 - y)^2\right)}  - (2 - y)
\right.\nonumber\\
&&\left.-\left(\slashed{\rho} - 1 \right)\left\{
-\frac{y}{m}  + 2 i s \frac{y(1-y)}{m} 
\left(2 - y\right)
\right\}\right].
\nonumber\\
\label{eq:renself0}
\end{eqnarray}

\section{Renormalized self-energy and mass renormalization at finite $B$}\label{SecIV}

From the definition of the counterterms in the previous section, in particular Eq.~(\ref{eq:renself0}), we have that the renormalized self-energy for any finite
strength of the magnetic field is given by
\begin{eqnarray}
\Sigma^{ren}(p,B) &=& \frac{2m e^2}{(4\pi)^2}\int_{0}^{\infty}\frac{ds}{s}\int_{0}^{1}dye^{i s \left( -(1-y)^2  + i\epsilon\right)}\nonumber\\
&\times&\left[
\left((A) + (B) - (C)\right)e^{i s \left(\varphi(y,\rho,B) + (1 - y)^2\right)}
\right.\nonumber\\
&-&(2 - y)-\left(\slashed{\rho} - 1 \right)\nonumber\\
&\times&\left.\left\{
-\frac{y}{m}  + 2 i s \frac{y(1-y)}{m} 
\left(2 - y\right)
\right\}\right],
\label{eq:renoselfB}
\end{eqnarray}
where the phase $\varphi(y,\rho,B)$ was defined in Eq.~(\ref{eq:phaseB}), while
the tensor coefficents $(A)$, $(B)$, and $(C)$ are given by Eqs.~(\ref{TermA})-~(\ref{TermC}). It is straightforward to verify that, by construction, the renormalized self-energy Eq.~(\ref{eq:renoselfB}) satisfies the renormalization conditions, Eq.~(\ref{eq:R1})
and Eq.~(\ref{eq:R2}), in the limit $B\rightarrow 0$, as they should.
The mass shift for a finite magnetic field strength is thus defined by
\begin{eqnarray}
\delta m_{B} = m_{B} - m &=& \left.\Sigma^{ren}(p,B)\right|_{\slashed{p}_{\parallel} = m}\nonumber\\
&=& \left.\Sigma^{ren}(p,B)\right|_{\slashed{\rho}_{\parallel} = 1},
\label{eq:massB1}
\end{eqnarray}
and, by construction, it clearly satisfies
\begin{eqnarray}
\lim_{B\rightarrow 0}\delta m_{B} = 0.
\label{eq:massBlim}
\end{eqnarray}

{Fixing the conditions $\slashed{\rho}_{\parallel}=1$ and $\slashed{\rho}_{\perp} = 0$ amounts to finding the particle’s energy in the lowest energy state, namely,
$p_0 = m_B$, which indicates that the three-vector components of the four
momentum are equal to zero. Since $p_0$ comes in combination with $p_3$ to form
$p_\parallel$ and this variable decouples from $p_\perp$, it is natural to first take
$p_\perp=0$ to later take $p_3$ equal to zero. Then,} as stated in Eq.~(\ref{eq:massB1}), we obtain the explicit integral expression for the magnetic mass shift operator
\begin{eqnarray}
\delta m_B &=& \frac{2m e^2}{(4\pi)^2}\int_{0}^{1}dy \int_{0}^{\infty}\frac{ds}{s} e^{i s \left( -(1-y)^2  + i\epsilon\right)}\nonumber\\
&\times&\left[
\frac{(2 - y)\cos(\mathcal{B}s(1-y))}{y\cos(\mathcal{B}s(1-y)) + \frac{\sin(\mathcal{B}s(1-y))}{\mathcal{B}s}}  - (2 - y)
\right.\nonumber\\
&-&\left.\frac{y \sin(\mathcal{B}s(1-y))i\gamma^1 \gamma^2\sgn(eB)}{y \cos(\mathcal{B}s(1-y)) - \frac{\sin(\mathcal{B}s(1-y))}{\mathcal{B}s} }\right].
\label{eq:deltamB2}
\end{eqnarray}
We notice that the physical origin of the third term in Eq.~(\ref{eq:deltamB2}) comes from the electromagnetic coupling $\frac{e}{2}\sigma^{\mu\nu}F^{\mu\nu} = e B i \gamma^{1}\gamma^{2}$, that determines a different value of the self-energy, and hence of the mass, for each spin component parallel or anti-parallel to the external magnetic field, respectively. In Appendix~\ref{Ap_Tsaisconnection}, we compare our result in Eq.~(\ref{eq:deltamB2}) with Ref.\cite{Tsai:1974df}.

It is convenient to express the operator in terms of the projectors
\begin{eqnarray}
\hat{O}^{(\pm)} = \frac{1}{2}\left(\mathbf{1}
\pm i\gamma^1\gamma^2 \sgn(eB)
\right),
\end{eqnarray}
such that we have
\begin{eqnarray}
\delta m_B = \hat{O}^{(+)}\delta m_B^{(+)} 
+ \hat{O}^{(-)}\delta m_B^{(-)}. 
\end{eqnarray}
Here, the magnetic mass shift components are given by
\begin{eqnarray}
\delta m_B^{(\pm)} &=& \frac{2m e^2}{(4\pi)^2}\int_{0}^{1}dy \int_{0}^{\infty}\frac{ds}{s} e^{i s \left( -(1-y)^2  + i\epsilon\right)}\nonumber\\
&\times&\left[
\frac{(2 - y)\cos(\mathcal{B}s(1-y))}{y\cos(\mathcal{B}s(1-y)) + \frac{\sin(\mathcal{B}s(1-y))}{\mathcal{B}s}}  - (2 - y)
\right.\nonumber\\
&&\left.\mp \frac{y \sin(\mathcal{B}s(1-y)) }{y \cos(\mathcal{B}s(1-y)) + \frac{\sin(\mathcal{B}s(1-y))}{\mathcal{B}s} }
\right].
\end{eqnarray}

\begin{figure}[t]
\centering
    \includegraphics[scale=0.62]{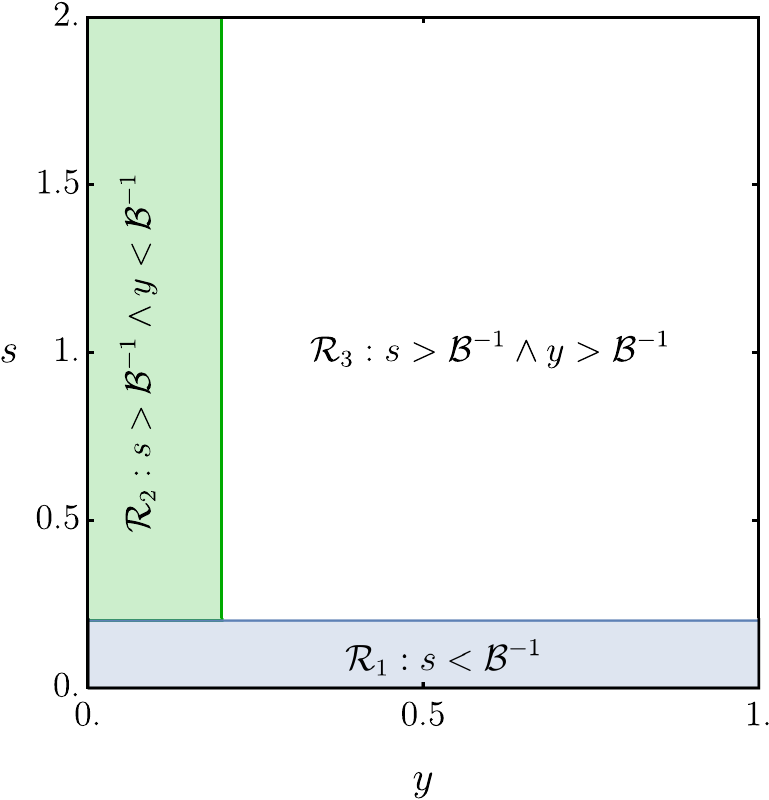}
    \caption{The three integration regions in the domain $(y,s)\in [0,1]\times[0,\infty]$, as described in the text.}
    \label{fig:intregions}
\end{figure}
% \begin{figure}[t]
% \centering
%     \includegraphics[width=0.5\textwidth]{QEDVertexB/figures/Integration_Regions.pdf}
%     \caption{The three integration regions in the domain $(y,s)\in [0,1]\times[0,\infty]$, as described in the text.}
%     \label{fig:intregions}
% \end{figure}

Notice that the presence of the counter-term makes the integrand to identically vanish in the limit $s\rightarrow 0$. Similarly, in the limit $y\rightarrow 0$, the integrand is finite. Therefore, for large magnetic fields, $\mathcal{B} \gg 1$,
it is convenient to split the integration domain $(y,s) \in [0,1]\times[0,\infty]$ into three separate regions $(\mathcal{R}_1,\mathcal{R}_2,\mathcal{R}_3)$, as depicted in 
Fig.~\ref{fig:intregions}, as follows
\begin{eqnarray}
&&\int_{0}^1 dy \int_{0}^{\infty}ds\\
&=&\underbrace{\int_{0}^1 dy\int_{0}^{\mathcal{B}^{-1}}}_{\mathcal{R}_1}ds
+\underbrace{\int_0^{\mathcal{B}^{-1}}dy\int_{\mathcal{B}^{-1}}^{\infty}}_{\mathcal{R}_2}ds+\underbrace{ \int_{\mathcal{B}^{-1}}^{1}dy\int_{\mathcal{B}^{-1}}^{\infty}}_{\mathcal{R}_3}ds\nn,
\label{separate}
\end{eqnarray}

such that we write the mass shift components as
\begin{eqnarray}
\delta m_B^{(\pm)} &=& \left.\delta m_B^{(\pm)}\right|_{\mathcal{R}_1} +
\left.\delta m_B^{(\pm)}\right|_{\mathcal{R}_2}+\left.\delta m_B^{(\pm)}\right|_{\mathcal{R}_3}.
\label{Eq:massshift}
\end{eqnarray}

The integrand in the first term, where $\mathcal{B}s < 1$, is bounded from above, and its singularity as $s\rightarrow 0$ is removed by the counter-term. Moreover (see Appendix~\ref{AppC} for details), its contribution is
{
\begin{eqnarray}
\left.\delta m_B^{(\pm)}\right|_{\mathcal{R}_1}&=& \frac{2 m e^2}{(4\pi)^2}
\left(-\frac{157}{2016} + \frac{2041}{56700}i\mathcal{B}^{-1}\right.\nonumber\\
&\mp&
\left.\left[ \frac{91}{540}
-\frac{257}{10080}i\mathcal{B}^{-1}
\right]\right) + O(\mathcal{B}^{-2}).
\end{eqnarray}}

Similarly, as shown in Appendix~\ref{AppC},
the contribution arising from the second region
is (for $\mathcal{B}\gg 1$)
{\begin{eqnarray}
\left.\delta m_B^{(\pm)}\right|_{\mathcal{R}_2}
&&\sim \frac{2 m e^2}{(4\pi)^2}
\left\{
-1 - 2\mathcal{B}^{-1}\ln(\mathcal{B})
+ \mathcal{B}^{-1}\left(2\gamma\right.\right.\\
&&\left.\left.- 2\ln\left[|1 - e^{2i} | \right]
+ i\left(\pi \pm \frac{1}{2} \right)\right)
\right\}
+ O(\mathcal{B}^{-2}).\nn
\end{eqnarray}}

The contribution of both of these subleading
kinematic
regions becomes finite and strictly real as $\mathcal{B}\rightarrow\infty$. 

Finally, for the remaining, and hence dominant, kinematic region, characterized by the condition $\mathcal{B}s \gg 1$, 
since $|\sin(\mathcal{B}s(1 - y))| \le 1$, we have the inequality
\begin{eqnarray}
\left|\frac{\sin(\mathcal{B}s(1 - y))}{\mathcal{B}s}\right| \le \frac{1}{\mathcal{B}s} \ll 1
\end{eqnarray}
Under this assumption, the asymptotic expression for the mass shift components in Eq.~(\ref{eq:deltamB2}) is
\begin{eqnarray}
\left.\delta m_B^{(\pm)}\right|_{\mathcal{R}_3} &=& \frac{2m e^2}{(4\pi)^2}\int_{\mathcal{B}^{-1}}^{1}dy \int_{\mathcal{B}^{-1}}^{\infty}\frac{ds}{s} \nonumber\\
&\times&e^{i s \left( -(1-y)^2  + i\epsilon\right)}\left[
\frac{(2 - y)(1 - y)}{y}\right.\nonumber\\
&&\left.
\mp  \tan(\mathcal{B}s(1-y)) 
\right].
\label{eq:deltamB3}
\end{eqnarray}
%{Since the $\tan(z)$ function has a %strongly oscillatory behavior, we use the %following geometric series expansion
%%\begin{eqnarray}
%\!\!\!\!\!\!\!\tan(\mathcal{B}s(1-y))\!&=&\! %-i\left(1 + 2 \sum_{n=1}^{\infty}(-1)^n e^{-i 2 %n \mathcal{B}s(1-y)}\right)\!\!.
%\end{eqnarray}}

We notice the presence of the $\tan(\mathcal{B}s(1-y))$ function
in the integrand of Eq.~(\ref{eq:deltamB3}),
which is continuous when defined
in the open interval $(-\frac{\pi}{2},\frac{\pi}{2})$, and
extends itself periodically along the real axis, namely,
$\tan(x + n\pi) = \tan(x)$, possessing infinitely many poles at every
odd multiple of $\pi/2$, {\it i.e.} for $x \rightarrow \pm\frac{(2n-1)\pi}{2}$,
$\tan(x)\to\pm\infty$. Therefore, the
integral in Eq.~(\ref{eq:deltamB3}) must
be interpreted as a principal value and hence, in the sense of distributions, we can use the periodic series (see Appendix~\ref{AppD} for details)
\begin{eqnarray}
\label{eq:tanseries}
\tan(&\!\!\mathcal{B}\!\!&s(1-y))=2 \sum_{n=1}^{\infty}(-1)^{n-1} \sin(2\mathcal{B}s(1-y))\nonumber\\
&=& i\sum_{n=1}^{\infty}(-1)^n \left( e^{2in\mathcal{B}s(1-y)}- e^{-2in\mathcal{B}s(1-y)}\right).
\end{eqnarray}

Substituting Eq.~(\ref{eq:tanseries}) into Eq.~(\ref{eq:deltamB3}),
and using the exact definitions (as $\epsilon \rightarrow 0^{+}$),
\begin{eqnarray}
\int_{\mathcal{B}^{-1}}^{\infty}\frac{ds}{s} e^{i s \left( -(1-y)^2  + i\epsilon\right)} &=&
\Gamma\left(0,i \frac{(1-y)^2}{\mathcal{B}} \right),
\end{eqnarray}
\begin{eqnarray}
&&\int_{1/\mathcal{B}}^{\infty}\frac{ds}{s} e^{i s \left( -(1-y)(1 - y \pm 2 n \mathcal{B})  + i\epsilon\right)}=\nonumber\\
&&\Gamma\left(0,i \frac{(1-y)(1 - y \pm 2 n \mathcal{B})}{\mathcal{B}} \right),
\end{eqnarray}
we obtain that the magnetic mass shift components are given by
%{
%\begin{eqnarray}
%&&\left.\delta %m_B^{(\pm)}\right|_{\mathcal{R}_3} = \frac{2m %e^2}{(4\pi)^2}\int_{\mathcal{B}^{-1}}^{1}\frac{%dy}{y} \Big[
%(2 - y)\nonumber\\
%&\times&
%\left.(1 - y)\Gamma\left(0,i %\frac{(1-y)^2}{\mathcal{B}} \right)
%\pm i\left\{
%\Gamma\left(0,i \frac{(1-y)^2}{\mathcal{B}} %\right)\right.\right.\nonumber\\
%&+&
%2\sum_{n=1}^{\infty}(-1)^n
%\Gamma\left(0,i \frac{(1-y)(1 - y + 2 n %\mathcal{B})}{\mathcal{B}} \right)\left.
%\Big\}
%\right].
%\label{eq:deltamB5}
%\end{eqnarray}
%}
\begin{eqnarray}
&&\left.\delta m_B^{(\pm)}\right|_{\mathcal{R}_3} = \frac{2m e^2}{(4\pi)^2}\int_{\mathcal{B}^{-1}}^{1}dy \Bigg[
\frac{(2 - y)(1 - y)}{y}\nonumber\\
&\times&
\Gamma\left(0,i \frac{(1-y)^2}{\mathcal{B}} \right)\nn\\
&&\mp i\sum_{n=1}^{\infty}(-1)^n\left\{
\Gamma\left(0,i \frac{(1-y)(1 - y - 2 n \mathcal{B})}{\mathcal{B}}\right)\right.\nn\\
&&\left.-\Gamma\left(0,i \frac{(1-y)(1 - y + 2 n \mathcal{B})}{\mathcal{B}}\right)
\right\}
\Bigg].
\label{eq:deltamB5}
\end{eqnarray}

We notice that the incomplete Gamma function $\Gamma(0,z)$ satisfies the following identity
\begin{eqnarray}
\Gamma(0,i z) = -\gamma - \ln(i z) - \sum_{k=1}^{\infty}\frac{(-i z)^k}{k\, (k!)}.
\end{eqnarray}

Therefore, for large $\mathcal{B}\gg 1$ we have the following asymptotics
\begin{eqnarray}
\!\!\!\!\!\!\!\!\Gamma\left(0,i \frac{(1-y)^2}{\mathcal{B}} \right) \!\!=\!\! -\gamma - 
\ln\left(i \frac{(1-y)^2}{\mathcal{B}}\right)\!+ O\left(\mathcal{B}^{-1}\right)\!,
\end{eqnarray}
thus leading to a double logarithmic dependence of the mass shift operator $\delta m_B \sim \left(\ln(\mathcal{B})\right)^2$ for $\mathcal{B}\gg 1$.
More precisely, as shown in Appendix~\ref{AppD},
the integrals involved in Eq.~(\ref{eq:deltamB5}) display the following asymptotic behavior (for $\mathcal{B}\gg1$),
\begin{eqnarray}
&&\int_{\mathcal{B}^{-1}}^{1}\frac{dy}{y}(1 - y)(2 - y)\Gamma[0,i\mathcal{B}^{-1}(1-y)^2]\nn\\
&=& 2\left[\ln(\mathcal{B})\right]^2 - \left( 2\gamma + \frac{5}{2} + i\pi\right)\ln(\mathcal{B}) +  O(\mathcal{B}^{0}),
\end{eqnarray}
%\begin{eqnarray}
%\int_{\mathcal{B}^{-1}}^{1}\frac{dy}{y}\Gamma[0,i\mathcal{B}^{-1}(1-y)^2] 
%&=& \left[\ln(\mathcal{B})\right]^2
%-\left(\gamma + \frac{i\pi}{2}\right)\ln(\mathcal{B})\nonumber\\
%&+& O(\mathcal{B}^{0}),
%\label{assymptotic2}
%\end{eqnarray}
and the infinite sum over Landau levels in Eq.~(\ref{eq:deltamB5}), as shown in Appendix~\ref{AppE},  is
given by
\begin{eqnarray}
&&\pm i\sum_{n=1}^{\infty}\int_{\mathcal{B}^{-1}}^{1}dy\left\{\Gamma[0,i\mathcal{B}^{-1}(1-y)(1 -y - 2 n \mathcal{B})]\right.\nn\\
&&\left.-
\Gamma[0,i\mathcal{B}^{-1}(1-y)(1 -y + 2 n \mathcal{B})]\right\}
\nn\\
&&= \mp 0.421794\left(1-\mathcal{B}^{-1} \right) \mp \ln(2) \pm i\left(1-\mathcal{B}^{-1}\right)\frac{\ln(2)}{2}\nn\\
&&+ O(\mathcal{B}^{-2}).
\label{eq:intgamman}
\end{eqnarray}

Therefore, the dominant contribution to the
magnetic mass shift components is, after Eq.~(\ref{eq:deltamB5}) (for $\mathcal{B}\gg 1$)
%{
%\begin{eqnarray}
%\left.\delta m_B^{(\pm)}\right|_{\mathcal{R}_3} %&=& \frac{2m e^2}{(4\pi)^2}
%\left\{
%\left(2 \pm i  %\right)\left[\ln(\mathcal{B})\right]^2\right.\n%n\\
%&&\left.-\left[2\gamma + \frac{5}{2} + i\pi \mp %i\left( -\gamma- %\frac{i\pi}{2}\right.\right.\right.\nn\\
%&&\left.\left.\left. + 0.81878 + 0.421645 i%\right.\Big)\right.\Big]\ln(\mathcal{B})\right%.\Big\}\nn\\ 
%&&+ O(\mathcal{B}^0).
%\end{eqnarray}
%}
\begin{eqnarray}
\left.\delta m_B^{(\pm)}\right|_{\mathcal{R}_3} &=& \frac{2m e^2}{(4\pi)^2}
\Bigg\{
2\left[\ln(\mathcal{B})\right]^2-\left[2\gamma + \frac{5}{2}  + i\pi
\right]\ln(\mathcal{B})\Bigg\}\nn\\
&&+ O(\mathcal{B}^0).
\label{R3}
\end{eqnarray}

{ Note that the above result is the same when only the LLL is taken into account. Moreover, given that Eq.~(\ref{R3}) is independent of the sign of the projection, we rename it as $\left.\delta m_B\right|_{\mathcal{R}_3}$.}

\begin{figure}
    \centering
    \!\!\includegraphics[scale=0.6]{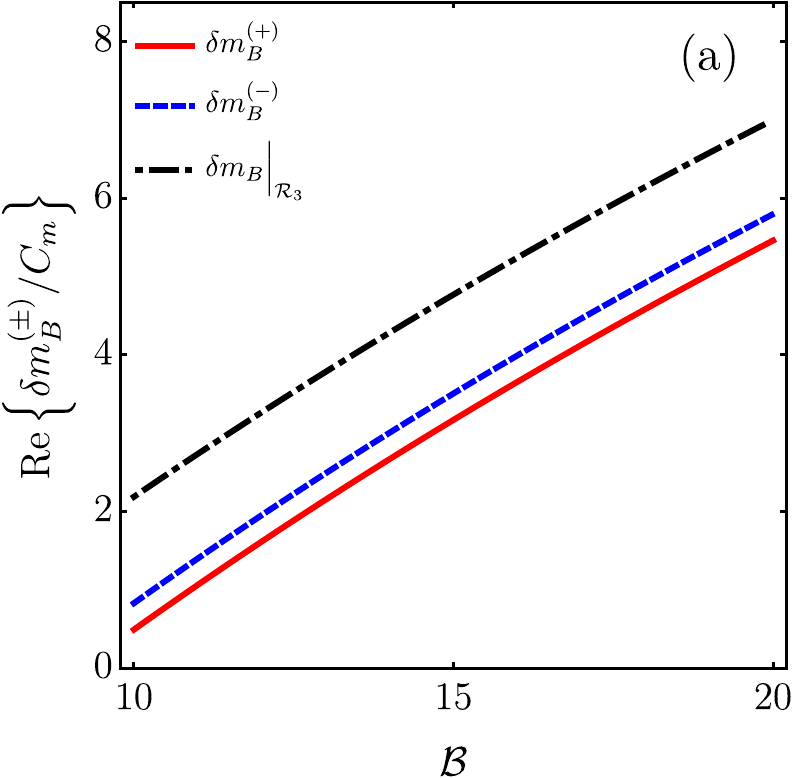}\nn\\
    \vspace{0.3cm}
    \includegraphics[scale=0.6]{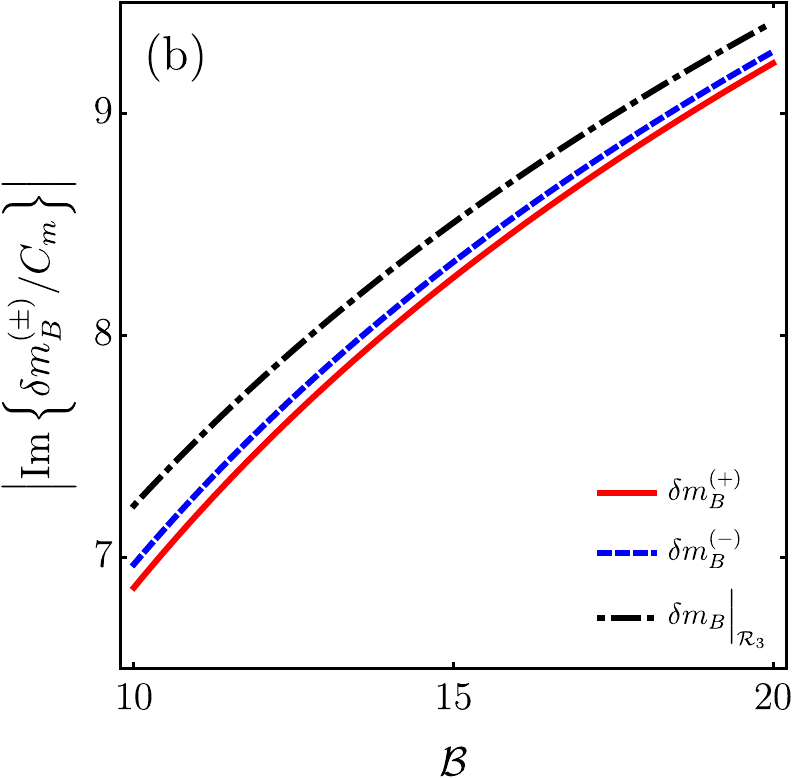}
    \caption{{(a) Real and (b) imaginary parts for both projections $(\pm)$ of the mass-shift given in Eq.~(\ref{Eq:massshift}), compared with the dominant $\left.\delta m_B\right|_{\mathcal{R}_3}$ contribution.}}
    \label{Fig:ReImAndLLL}
\end{figure}

\begin{figure}
    \centering
    \!\!\includegraphics[scale=0.6]{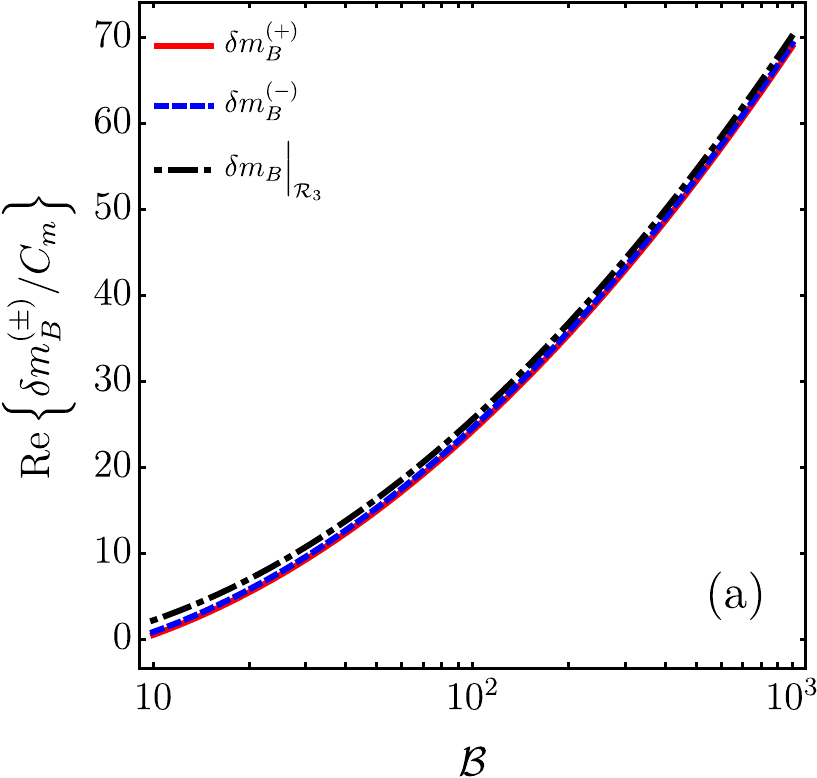}\nn\\
    \vspace{0.3cm}
    \includegraphics[scale=0.6]{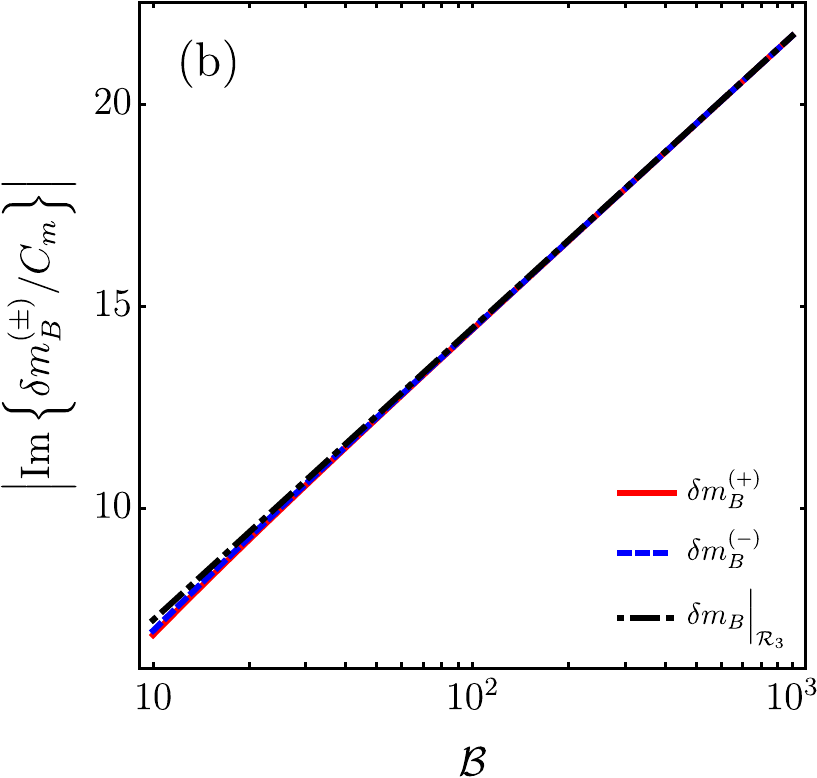}
    \caption{{(a) Real and (b) imaginary parts for both projections $(\pm)$ of the mass-shift given in Eq.~(\ref{Eq:massshift}), compared with the dominant $\left.\delta m_B\right|_{\mathcal{R}_3}$ contribution.}}
    \label{Fig:ReImAndLLL_log}
\end{figure}

\section{Analysis of the results}\label{SecV}

In order to interpret the results obtained, let us write the inverse
propagator in momentum space and at finite magnetic field in the form
\begin{eqnarray}
 \left.[-i S_F(p)\right]^{-1} &=& \slashed{p} - m - \Sigma(p,B)\\
&=& \left( \hat{O}^{(+)} + \hat{O}^{(-)} \right)\left(\slashed{p} - m\right) - \hat{O}^{(+)}\Sigma^{(+)}(p,B)\nn\\
&-& \hat{O}^{(-)}\Sigma^{(-)}(p,B)\nn\\
&=& \hat{O}^{(+)} \left[-i \Delta_F^{(+)}(p)\right]^{-1}
+ \hat{O}^{(-)} \left[-i \Delta_F^{(-)}(p)\right]^{-1}\nn
\end{eqnarray}
where the Feynman propagators for each spin projection $(\pm)$ parallel to the magnetic field direction are thus given by
\begin{eqnarray}
\Delta_F^{(\pm)}(p) = \frac{i}{\slashed{p} - m - \Sigma^{(\pm)}(p,B) + i\epsilon}.
\end{eqnarray}

As obtained after the explicit calculations in the previous section,
the pole in each of these propagators contains an imaginary part, whose magnitude
scales as $\Im \Sigma^{(\pm)}(m,B)\sim -\ln(\mathcal{B})$, while the real part scales as $\Re \Sigma^{(\pm)}(m,B)\sim \left[\ln(\mathcal{B})\right]^2$, and hence the latter becomes dominant at large magnetic fields, $\mathcal{B}\gg 1$. Therefore, the physical mass is determined by the real part,
\begin{eqnarray}
m_B^{(\pm)} = m + \Re \Sigma^{(\pm)}(m,B),
\end{eqnarray}
while the imaginary part determines an spectral width, since near the pole
\begin{eqnarray}
\!\!\!\!\!\!\!\!\!\Delta_F^{(\pm)}(p) &\sim& \frac{i}{\slashed{p} - m_{B}^{(\pm)} - i\Im \Sigma^{(\pm)}(m,B) + i\epsilon}\nn\\
&\sim& i\frac{\slashed{p} + m_{B}^{(\pm)} + i\Im \Sigma^{(\pm)}(m,B)}{p^2 - \left(m_{B}^{(\pm)}\right)^2 - 2 i m_{B}^{(\pm)}\Im \Sigma^{(\pm)}(m,B)}.
\end{eqnarray}

Figures~\ref{Fig:ReImAndLLL} and~\ref{Fig:ReImAndLLL_log} show the behavior of the real and imaginary parts for both projections $(\pm)$ of the mass-shift given in Eq.~(\ref{Eq:massshift}), compared with the dominant{ $\left.\delta m_B\right|_{\mathcal{R}_3}$. All the contributions are normalized to the quantity $C_m=2me^2/(4\pi)^2$}. Figure~\ref{Fig:ReImAndLLL} shows the case for a moderate range of the field strength in units of the fermion mass whereas Fig.~\ref{Fig:ReImAndLLL_log} is for the case of a larger range of field strengths. { It is important to mention, that our approximation is valid in the region where $\ln\mathcal{B}\geq 1$, therefore, the results have $\mathcal{B}\sim 10$ as a lower limit.} Notice that as the field strength increases, both the real and the imaginary parts are better described by the dominant contribution which in turn comes from the LLL { or $\left.\delta m_B\right|_{\mathcal{R}_3}$}.

{ The appearance of an imaginary part is an interesting and expected feature, which has a direct physical picture: An electron not yet affected by the
magnetic field (corresponding to the external leg in the self-energy diagram)
enters a region where the magnetic field forces it to occupy a Landau level and
thus (in classical terms), an orbit around the field lines. The emission of photons comes from the fact that the fermion need to preserve its energy and momentum (a syncroton-like process). At lowest order, this is represented by the emission of one photon, but as we shall see, when the Landau levels are close to each other, the process
is better represented by a distribution of Landau levels that can possibly be
occupied and thus the need of the spectral function representation. In that spirit, note that} the real and imaginary parts combined define a Breit-Wigner resonance $\Gamma^{(\pm)}= -2 \Im \Sigma^{(\pm)}(m,B)$, whose relative width
\begin{eqnarray}
\frac{\Gamma^{(\pm)}}{m_{B}^{(\pm)}} &=& -\frac{2 \Im \Sigma^{(\pm)}(m,B)}{m_{B}^{(\pm)}}\sim \frac{\ln(\mathcal{B})}{\left[ \ln(\mathcal{B}) \right]^2} \sim \left[ \ln(\mathcal{B}) \right]^{-1}
\end{eqnarray}
decreases to zero at a rate $\left[ \ln(\mathcal{B}) \right]^{-1}$ as
the magnetic field grows large ($\mathcal{B}\rightarrow\infty$). This effect can be highlighted analyzing the spectral density,
which is, as usual, calculated from the imaginary part of the scalar denominator in the Feynman propagator. For a well defined single-particle state with mass $m_{B}^{(\pm)}$, we would have (as $\epsilon\rightarrow 0^{+}$)
\begin{eqnarray}
\widetilde{\rho}(p^2) &=& -\frac{1}{\pi}\Im\left( \frac{1}{p^2 -  \left(m_{B}^{(\pm)}\right)^2  + i\epsilon} \right)\nn\\
&=& \frac{\epsilon/\pi}{\left(p^2 - \left( m_{B}^{(\pm)} \right)^2\right)^2 + \epsilon^2}\nn\\
&\xrightarrow[\epsilon\to 0^{+}]&\,\,\, \delta\left[p^2 - \left( m_{B}^{(\pm)} \right)^2\right].
\end{eqnarray}

In the present case, however, due to the presence of a finite imaginary part in the pole,
a similar calculation leads to a spectral density of the form
\begin{eqnarray}
\widetilde{\rho}(p^2) &=& -\frac{1}{\pi}\Im\left( \frac{1}{p^2 -  \left(m_{B}^{(\pm)}\right)^2  + i m_{B}^{(\pm)}\Gamma^{(\pm)} + i\epsilon} \right)\nn\\
&=& \frac{\epsilon/\pi + m_{B}^{(\pm)}\Gamma^{(\pm)}/\pi}{\left(p^2 - \left( m_{B}^{(\pm)} \right)^2\right)^2 + \left(m_{B}^{(\pm)}\Gamma^{(\pm)} + \epsilon\right)^2}\\
&\sim& 
%\delta(p^2 - \left( m_{B}^{(\pm)} \right)^2) + 
\frac{ m_{B}^{(\pm)}\Gamma^{(\pm)}/\pi}{\left(p^2 - \left( m_{B}^{(\pm)} \right)^2\right)^2 + \left[m_{B}^{(\pm)}\Gamma^{(\pm)}\right]^2 },
\nn
\label{Eq:rho}
\end{eqnarray}
that clearly shows %a stable single-particle
%like resonance with a well defined mass,
%followed by 
a smeared, roughly Lorentzian
distribution, representing a quasi-continuum of unstable energy states. While the
dominant contribution at very large values of the magnetic field can be easily traced back to the LLL,
the smearing is related to
the probability to populate the higher
Landau levels, with a relative width
$\Gamma^{(\pm)}/m_{B}^{(\pm)} \sim \left[\ln(\mathcal{B})\right]^{-1}$
that decays to zero as $\mathcal{B}\rightarrow\infty$, and hence
in this limit all the spectral weight
is concentrated on the stable LLL.
\begin{figure}
    \centering
    \!\!\includegraphics[scale=0.6]{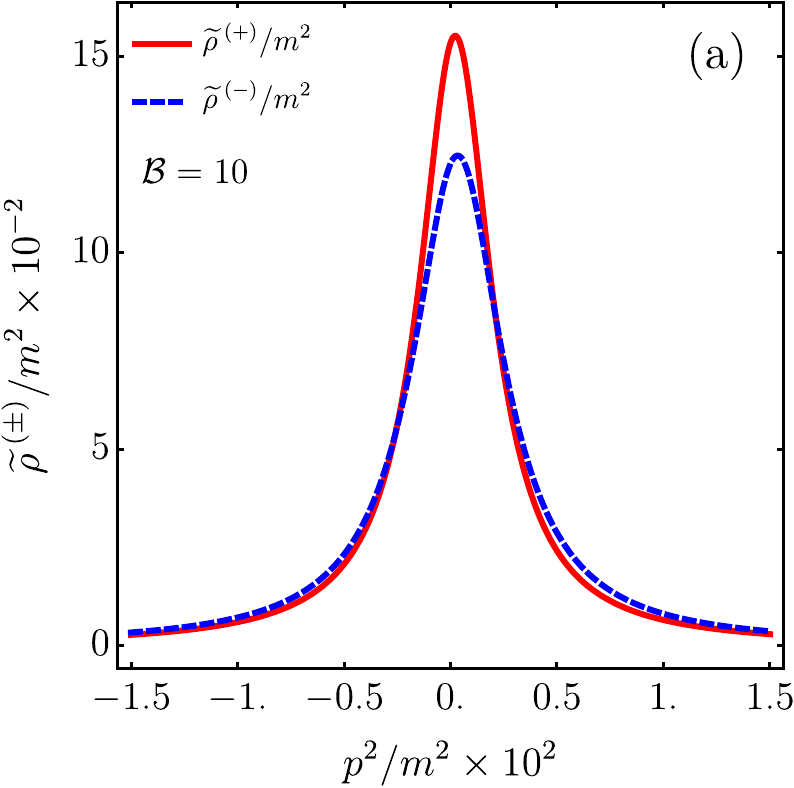}\nn\\
    \vspace{0.3cm}
    \includegraphics[scale=0.6]{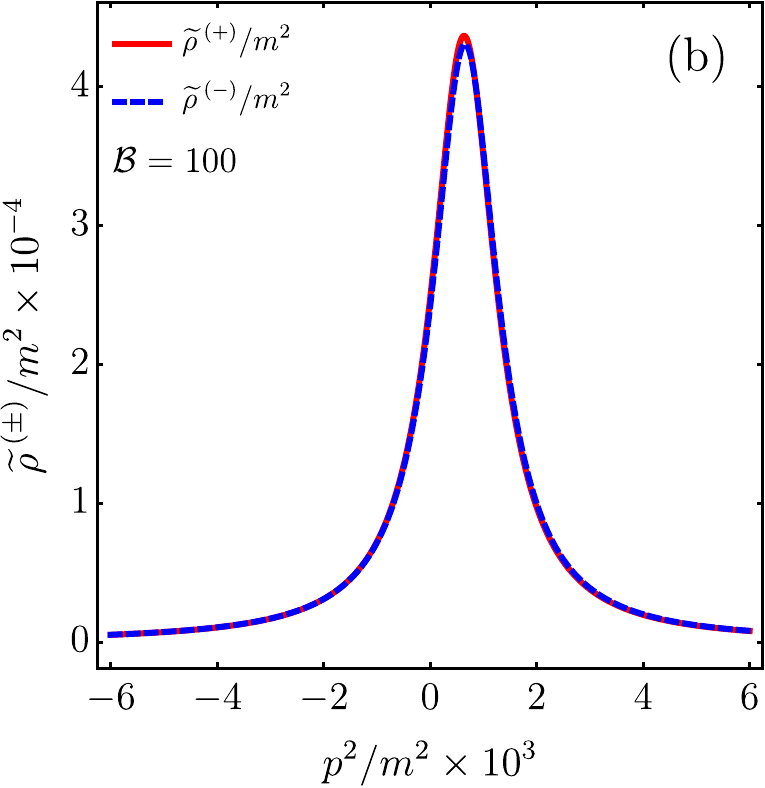}
    \caption{Spectral density from Eq.~(\ref{Eq:rho}) for (a) $\mathcal{B}=10$ and (b) $\mathcal{B}=100$ as a function of the momentum squared. The solid line represents the contribution of $\delta m_B^{(+)}$ whereas the dashed line is the contribution of $\delta m_B^{(-)}$.}
    \label{Fig:Spectral_Density}
\end{figure}
Figure~\ref{Fig:Spectral_Density}
shows the spectral densities as functions of the momentum squared scaled by the fermion mass squared, for two values of the magnetic field $\mathcal{B}=10^2$ (a) and  $\mathcal{B}=10^3$ (b). Notice that as the field strength increases, the peaks for both modes approach each other and at the same time, the Lorentzian distribution narrows.

\section{Conclusions}\label{Concl}

In summary, we have studied the fermion self-energy in QED in the presence of a magnetic field. After carrying out the renormalization procedure we have shown that in the large field limit, when accounting for sub-leading contributions associated to Landau levels other than the LLL, the mass function develops also an imaginary part. From this imaginary part it is possible to define a spectral density describing the spread of the mass function in momentum which is centered at the magnetic-field modified mass. The width of this distribution becomes small as the field intensity increases in such a way that for asymptotic values of the field, when the separation between Landau levels becomes also large, the mass function describes a stable particle occupying only the LLL. For large but finite values of the magnetic field, the spectral density represents the finite probability for the fermion to occupy Landau levels other than the LLL.

{ The present calculation has potential applications. Recent works consider dilepton production from a single photon in a strong magnetic field, giving rise to vacuum dichroism, i.e., the spectrum becomes anisotropic with respect to the magnetic field direction, depending also on the photon polarization.~\cite{hattori2021di}. There, the usual QED vertex is considered when the outgoing fermions are dressed by corrections due to the external magnetic field. Nevertheless, self-energy correction for these outgoing fermions is not taken into account, which would soft the spiky structure of the spectrum through the imaginary part that we report. On the other hand, in the field of condensed matter systems, the vacuum-polarization diagram plays an important role in the optical conductivity and transparency in graphene~\cite{PhysRevB.98.195430,valenzuela2015graphene}. Normally, the fermion propagators are dressed by external field corrections but the inclusion of quantum self-energy corrections for the fermion propagators, as we have done in this paper, have not been considered. Certainly it would be interesting to explore the relevance of this correction for the optical transparency of graphene in the presence of an external magnetic field. 

Finally, it is important to mention the range of validity of our approach, in terms of the magnetic field strength. As discussed in the main text, we focus in the region where $\ln\mathcal{B}\geq 1$ so that $\mathcal{B}\sim 10$ is a lower limit of validity, and no restrictions over the upper limit were done. However, notice that, when the real part of the magnetic field-dependent coefficient that multiplies the mass, $\eta$, that represents the dimensionless mass correction, becomes of order $\mathcal{O}(1)$, it is important to re-sum all the leading double logarithmic corrections. As it is discussed in Ref.~\cite{Gusynin:1998nh} when the resummation becomes singular at some value of $\eta$, this signals the breakdown of perturbation theory and the transition to the non-perturbative regime is heralded. In that regime, the coupling may also receive important magnetic field corrections which certainly deserve a thorough investigation but are outside the scope of the present work.}

\section*{Acknowledgements}

This work was supported in part by UNAM-DGAPA-PAPIIT grant number IG100219 and by Consejo Nacional de Ciencia y Tecnolog\'{\i}a grant numbers A1-S-7655 and A1-S-16215. E. M. acknowledges support from FONDECYT (Chile) under grant No.
1190361, and ANID PIA/Anillo Grant No. ACT192023. M. Loewe acknowledges support from FONDECYT (Chile) under grants No. 1170107, 1190192 and from Conicyt/PIA/Basal (Chile) grant number FB0821.

%\section{Introduction}
%\label{sec:Intro}

%\emph{Introduction.}  
%%%%%%%%%%%%%%%%%%%%%%%%%%%%%%%%%%%%%%%%%%%%%%%%%%%%%%%%%%%%
%\emph{Acknowledgment.} .
\appendix

%---------------------------------
\section{Gauging away the phase factor}\label{AppA}
It is well known that in the presence of an external magnetic field, charged particle propagators can develop a phase factor $\Phi(x,x')$. In particular, the fermion propagator $S_f(x,x')$ including the phase factor, is  given by
\bea
S_F(x,x')=\Phi(x,x')\int\frac{d^4p}{(2\pi)^4}e^{-p\cdot(x-x')}S_F(p),
\eea
where $S_F(p)$ is given by Eq.~(\ref{Eq:Invariant_propagator}) and the phase takes the form
\bea
\Phi(x,x')=\exp\Big\{ie\int_x^{x'}d\xi^\mu\left[A_\mu+\frac{1}{2}F_{\mu\nu}(\xi-x')^\nu\right]\Big\}.\nn\\
\eea

In order to perform the integral, let us take a straight line path parametrized as
\bea
\xi^\mu=x'^\mu+t(x^\mu-x'^\mu),\text{ for }0< t <1.
\eea
Therefore, the phase actor becomes
\bea
\Phi(x,x')=\exp\left[ie\int_0^1 A_\mu(x^\mu-x'^\mu)dt\right],
\eea
where the anti-symmetry of $F_{\mu\nu}$ was used.

From the above, the phase can be removed by the gauge transformation
\bea
A_\mu(\xi)\to A'_\mu(\xi)+\frac{\partial}{\partial\xi^\mu}\alpha(\xi).
\eea

For our case, where the magnetic field is oriented in the $z$-direction, we have
\begin{subequations}
\bea
A_\mu=\frac{B}{2}(0,-x_2,x_1,0).
\eea
Therefore, choosing
\bea
\alpha(\xi)=\frac{B}{2}(x_2'\xi^1-x_1'\xi_2'),
\eea
\end{subequations}
we obtain
\bea
A_\mu'=\frac{B}{2}(0,x_2-x_2',x_1-x_1',0),
\eea
which implies that
\bea
A_\mu'(x^\mu-x'^\mu)=0,
\eea
and thus the phase can be safely removed. 
%---------------------------------

\section{Dirac algebra}\label{AppB}
Here we list the properties of products of Dirac matrices that appear in the calculation of the fermion self-energy in the presence of a magnetic field:
\begin{eqnarray}
\gamma_{\mu}\gamma^{\mu} &=& 4\\
\gamma_{\mu}\gamma^{1}\gamma^{2}\gamma^{\mu} &=& 4 g^{12} = 0\\
\gamma^{\mu}\slashed{k}_{\parallel}\gamma_{\mu} &=& k_{\parallel,\nu}\gamma^{\mu}\gamma^{\nu}\gamma_{\mu} = -2 k_{\parallel,\nu} \gamma^{\nu} = -2 \slashed{k}_{\parallel}\\
\gamma^{\mu}\slashed{k}_{\perp}\gamma_{\mu} &=& -2 \slashed{k}_{\perp}\\
\gamma_{\mu}\gamma^{1}\gamma^{2}\slashed{k}_{\parallel}\gamma^{\mu} &=& k_{\parallel,\nu}\gamma^{\mu}\gamma^{1}\gamma^{2}
\gamma^{\nu}\gamma_{\mu} = -2 \slashed{k}_{\parallel}\gamma^{1}\gamma^{2}
\end{eqnarray}

%------------------------

\section{Subdominant integration region}\label{AppC}
As presented in the main text, the magnetic mass shift components are given by 
\begin{eqnarray}
\delta m_B = \hat{O}^{(+)}\delta m_B^{(+)} 
+ \hat{O}^{(-)}\delta m_B^{(-)}, 
\end{eqnarray}
where the magnetic mass shift components are given by
the integral expressions
\begin{eqnarray}
&&\delta m_B^{(\pm)} = \frac{2m e^2}{(4\pi)^2}\nn\\
&\times&\int_{0}^{1}dy \int_{0}^{\infty}\frac{ds}{s}\,e^{i s \left( -(1-y)^2  + i\epsilon\right)}\nn\\
&\times&\left[
\frac{(2 - y)\cos(\mathcal{B}s(1-y))}{y\cos(\mathcal{B}s(1-y)) + \frac{\sin(\mathcal{B}s(1-y))}{\mathcal{B}s}}  - (2 - y)
\right.\nonumber\\
&&\left.\mp \frac{y \sin(\mathcal{B}s(1-y)) }{y \cos(\mathcal{B}s(1-y)) + \frac{\sin(\mathcal{B}s(1-y))}{\mathcal{B}s}}\right].
\end{eqnarray}

Moreover, as explained in the main text, we shall split the integration domain into three subregions, as follows
\begin{eqnarray}
\delta m_B^{(\pm)} &=& \left.\delta m_B^{(\pm)}\right|_{\mathcal{R}_1} +
\left.\delta m_B^{(\pm)}\right|_{\mathcal{R}_2}+\left.\delta m_B^{(\pm)}\right|_{\mathcal{R}_3}.
\end{eqnarray}

Let us now restrict ourselves to the kinematic region $0\le \mathcal{B}s < 1$, which corresponds to the $s$-integral domain
$s\in [0,\mathcal{B}^{-1})$. Within this region, we can Taylor expand the expressions in the square bracket as a power series in $\mathcal{B}s < 1$, to obtain
\begin{eqnarray}
&&\frac{(2 - y)\cos(\mathcal{B}s(1-y))}{y\cos(\mathcal{B}s(1-y)) + \frac{\sin(\mathcal{B}s(1-y))}{\mathcal{B}s}}  - (2 - y)\nn\\
&=&-\frac{ (2 - y)(1 - y)^3}{3}\left( \mathcal{B}s \right)^3
\nonumber\\
&-&\frac{(2 - y)(1 + 5 y)(1 - y)^5}{45}\left( \mathcal{B}s \right)^4 + O\left( \mathcal{B}s \right)^5,
\end{eqnarray}
and similarly
\begin{eqnarray}
&&\frac{y \sin(\mathcal{B}s(1-y)) }{y \cos(\mathcal{B}s(1-y)) + \frac{\sin(\mathcal{B}s(1-y))}{\mathcal{B}s} }
= y(1 - y)\left( \mathcal{B}s \right)\nn\\
&+&\frac{y^2(1-y)^3 }{3}\left( \mathcal{B}s \right)^3 + O\left( \mathcal{B}s \right)^5.
\end{eqnarray}
Integrating the first group of terms, we obtain
\begin{eqnarray}
&&\int_{0}^{1}dy \int_{0}^{\mathcal{B}^{-1}}\frac{ds}{s} e^{i s \left( -(1-y)^2  + i\epsilon\right)}\nn\\
&\times&\left[ 
\frac{(2 - y)\cos(\mathcal{B}s(1-y))}{y\cos(\mathcal{B}s(1-y)) + \frac{\sin(\mathcal{B}s(1-y))}{\mathcal{B}s}}  - (2 - y)
\right]\nn\\
&\sim& 
\frac{i \mathcal{B}^2}{360}\left[2 \left(30 i \left(2 \mathcal{B}^2+1\right) \text{Ei}\left(-\frac{i}{\mathcal{B}}\right)-14 i e^{-\frac{i}{\mathcal{B}}} \mathcal{B}^2\right.\right.\nn\\
&+&\left.\left.14 i \mathcal{B}^2-30 \pi  \mathcal{B}^2-60 i \gamma  \mathcal{B}^2+30 i \left(2 \mathcal{B}^2+1\right) \log (\mathcal{B})\right.\right.\nonumber\\
&&\left.\left.-7 e^{-\frac{i}{\mathcal{B}}} \mathcal{B}-24 \mathcal{B}-64 i e^{-\frac{i}{\mathcal{B}}}-30 i-15 \pi -30 i \gamma \right)\right.\nn\\
&+&\left.15 \sqrt[4]{-1} \sqrt{\pi } (\mathcal{B}-6 i) \sqrt{\mathcal{B}}\, \text{Erfi}\left(\frac{(-1)^{3/4}}{\sqrt{\mathcal{B}}}\right)\right]\nonumber\\
&=& -\frac{157}{2016} +\frac{2041 }{56700 }i \mathcal{B}^{-1} + \frac{14749}{1425600 }\mathcal{B}^{-2} + O(\mathcal{B}^{-3}),\nn\\
\end{eqnarray}
and similarly, for the second term
\begin{eqnarray}
&&\int_{0}^{1}dy \int_{0}^{\mathcal{B}^{-1}}\frac{ds}{s} e^{i s \left( -(1-y)^2  + i\epsilon\right)}\nn\\
&\times&\left[\frac{y \sin(\mathcal{B}s(1-y)) }{y \cos(\mathcal{B}s(1-y)) + \frac{\sin(\mathcal{B}s(1-y))}{\mathcal{B}s} }
\right]\nn\\
&\sim& 
\frac{1}{12} i \mathcal{B} \left[\left(6-4 \mathcal{B}^2\right) \text{Ei}\left(-\frac{i}{\mathcal{B}}\right)+2 \left(3-2 \mathcal{B}^2\right) \log (\mathcal{B})\right.\nn\\
&&\left.-6 \sqrt[4]{-1} \sqrt{\pi } \sqrt{\mathcal{B}} (\mathcal{B}-i) \text{Erf}\left(\frac{\sqrt[4]{-1}}{\sqrt{\mathcal{B}}}\right)+2 \mathcal{B} \left(\mathcal{B} \left(-3 e^{-\frac{i}{\mathcal{B}}}\right.\right.\right.\nn\\
&&\left.\left.\left.+3+2 \gamma -i \pi \right)+i\right)+12-6 \gamma +3 i \pi \right]
\nonumber\\
&&=\frac{91}{540}-\frac{257 }{10080}i\mathcal{B}^{-1} -\frac{307}{75600}\mathcal{B}^{-2} + O(\mathcal{B}^{-3}).
\end{eqnarray}
Combining both expressions, we obtain for the contribution of this first kinematic region
\begin{eqnarray}
\left.\delta m_B^{(\pm)}\right|_{\mathcal{R}_1}&=& \frac{2 m e^2}{(4\pi)^2}
\left(-\frac{157}{2016} + \frac{2041}{56700}i\mathcal{B}^{-1}\right.\nonumber\\
&\mp&
\left.\left[ \frac{91}{540}
-\frac{257}{10080}i\mathcal{B}^{-1}
\right]\right) + O(\mathcal{B}^{-2}).
\end{eqnarray}

Let us now consider the second integration region defined in the main text (see Fig.~\ref{fig:intregions}), corresponding to
$s\in (\mathcal{B}^{-1},\infty)$ and $y\in[0,\mathcal{B}^{-1})$. The last condition means that, for large magnetic fields $\mathcal{B} \gg 1$ the integration variable $y\ll 1$ within this interval. Therefore, a Taylor expansion around $y=0$ yields for the first term,
\begin{eqnarray}
&&e^{i s \left( -(1-y)^2  + i\epsilon\right)}(2 - y)\nn\\
&&\times\left[ 
\frac{\cos(\mathcal{B}s(1-y))}{y\cos(\mathcal{B}s(1-y)) + \frac{\sin(\mathcal{B}s(1-y))}{\mathcal{B}s}}  - 1
\right] = e^{-i s (1 - i\epsilon)}\left(\mathcal{B}s -\right.\nn\\
&&\left.\cot(\mathcal{B}s) \right) +
y e^{-i s (1 - i\epsilon)}\left(1 + 2 (\mathcal{B} s)^2 - 4 i s - \mathcal{B} s \cot(\mathcal{B} s)\right.\nn\\
&&\left.+ 4 i \mathcal{B} s^2 \cot(\mathcal{B} s)\right) + O(y^2),
\label{eq:c1}
\end{eqnarray}
and similarly for the second term
\begin{eqnarray}
&&e^{i s \left( -(1-y)^2  + i\epsilon\right)}\frac{y \sin(\mathcal{B}s(1-y)) }{y \cos(\mathcal{B}s(1-y)) + \frac{\sin(\mathcal{B}s(1-y))}{\mathcal{B}s} }\nn\\
&=& y e^{-i s (1 - i\epsilon)} \mathcal{B}s
 + O(y^2).
\label{eq:c2}
\end{eqnarray}
Integrating these expressions in their
corresponding kinematic region, we obtain
for Eq.~(\ref{eq:c1})
\begin{eqnarray}
&&\int_{0}^{\mathcal{B}^{-1}}dy\int_{\mathcal{B}^{-1}}^{\infty}\frac{ds}{s}e^{i s \left( -(1-y)^2  + i\epsilon\right)}(2 - y)\nn\\
&&\times\left[ 
\frac{\cos(\mathcal{B}s(1-y))}{y\cos(\mathcal{B}s(1-y)) + \frac{\sin(\mathcal{B}s(1-y))}{\mathcal{B}s}}  - 1
\right]\nn\\
&&\sim \int_{\mathcal{B}^{-1}}^{\infty}ds e^{-i s \left( 1  - i\epsilon\right)}
\left[ s - 2 i \mathcal{B}^{-2} - 2\frac{\mathcal{B}^{-1}}{s}
+ \frac{\mathcal{B}^{-2}}{2s}\right.\nn\\
&&\left.+\left(2 - \frac{\mathcal{B}^{-1}}{2} + 2 i  \mathcal{B}^{-1} s\right)\cot(\mathcal{B} s)\right]\nn\\
&&= \left(\frac{1}{2}\mathcal{B}^{-2} - 
2\mathcal{B}^{-1}\right)\Gamma[0,i\mathcal{B}^{-1}]
-e^{-i\mathcal{B}^{-1}} \left(1 + i\mathcal{B}^{-1} + 2\mathcal{B}^{-2}\right)\nn\\
&&+\left(2 - \frac{\mathcal{B}^{-1}}{2}  \right)
\left[
\frac{e^{i\left(2 - \mathcal{B}^{-1} \right)}}{2\mathcal{B}-1} {}_{2}F_{1}\left(1,1-\frac{1}{2\mathcal{B}};2-\frac{1}{2\mathcal{B}};e^{2i} \right)\right.\nn\\
&&\left.+ \frac{e^{i\left(-2 - \mathcal{B}^{-1} \right)}}{2\mathcal{B}+1} {}_{2}F_{1}\left(1,1+\frac{1}{2\mathcal{B}};2+\frac{1}{2\mathcal{B}};e^{-2i} \right)\right]\nn\\
&&+2\frac{i e^{i\left(2 - \mathcal{B}^{-1} \right)}}{\mathcal{B}^2\left(2\mathcal{B}-1\right)}{}_{2}F_{1}\left(1,1-\frac{1}{2\mathcal{B}};2-\frac{1}{2\mathcal{B}};e^{2i} \right)\nn\\
&&+ 2\frac{i e^{i\left(-2 - \mathcal{B}^{-1} \right)}}{\mathcal{B}^2\left(2\mathcal{B}+1\right)}{}_{2}F_{1}\left(1,1+\frac{1}{2\mathcal{B}};2+\frac{1}{2\mathcal{B}};e^{-2i} \right)\nn\\
&&+\frac{1}{2\mathcal{B}^3}\left\{
e^{-i \left(\frac{1}{\mathcal{B}}+2\right)} \Phi \left(e^{-2 i},2,1+\frac{1}{2 \mathcal{B}}\right)\right.\nn\\
&&-\left.e^{i \left(-\frac{1}{\mathcal{B}}+2\right)} \Phi \left(e^{-2 i},2,1-\frac{1}{2 \mathcal{B}}\right)
\right\}\nonumber\\
&&\sim -1 - 2\mathcal{B}^{-1}\ln(\mathcal{B})
+ \mathcal{B}^{-1}\left(2\gamma - 2\ln\left[|1 - e^{2i} | \right]  + i\pi\right)\nn\\ 
&&+ O(\mathcal{B}^{-2})
\end{eqnarray}
and for Eq.~(\ref{eq:c2})
\begin{eqnarray}
&&\int_{0}^{\mathcal{B}^{-1}}dy\int_{\mathcal{B}^{-1}}^{\infty}\frac{ds}{s}e^{i s \left( -(1-y)^2  + i\epsilon\right)}\nn\\
&&\times\frac{y \sin(\mathcal{B}s(1-y)) }{y \cos(\mathcal{B}s(1-y)) + \frac{\sin(\mathcal{B}s(1-y))}{\mathcal{B}s}}\nn\\
&&\sim -\frac{i}{2\mathcal{B}}e^{-i\mathcal{B}^{-1}}\\
&&= -\frac{i}{2}\mathcal{B}^{-1} + O(\mathcal{B}^{-2})\nn
\end{eqnarray}
In these expressions, we have used the periodic series expansion (see Appendix~\ref{AppD})
\begin{eqnarray}
\cot(\mathcal{B}s) = -i\sum_{n=1}^{\infty}\left(e^{2 i n \mathcal{B}s}-e^{-2 i n \mathcal{B}s} \right)
\end{eqnarray}
to obtain the analytical integrals
\begin{eqnarray}
&&\int_{\mathcal{B}^{-1}}^{\infty}ds e^{-i s \left( 1  - i\epsilon\right)}\cot(\mathcal{B}s)
=  \sum_{n=1}^{\infty}\left[
\frac{e^{-i\left(1-2 n\mathcal{B}  \right)\mathcal{B}^{-1}}}{-1 + 2 n \mathcal{B}}\right.\nn\\
&&\left.+ \frac{e^{-i\left(1+2 n\mathcal{B}  \right)\mathcal{B}^{-1}}}{1 + 2 n \mathcal{B}}
\right]\nn\\
&&=\frac{e^{i\left(2 - \mathcal{B}^{-1} \right)}}{2\mathcal{B}-1}{}_{2}F_{1}\left(1,1-\frac{1}{2\mathcal{B}};2-\frac{1}{2\mathcal{B}};e^{2i} \right)\nn\\
&&+ \frac{e^{i\left(-2 - \mathcal{B}^{-1} \right)}}{2\mathcal{B}+1}{}_{2}F_{1}\left(1,1+\frac{1}{2\mathcal{B}};2+\frac{1}{2\mathcal{B}};e^{-2i} \right)
\end{eqnarray}
and
\begin{eqnarray}
&&\int_{\mathcal{B}^{-1}}^{\infty}ds e^{-i s \left( 1  - i\epsilon\right)}s \cot(\mathcal{B}s)
= i\sum_{n=1}^{\infty}
\left[
\frac{e^{-i\left(1 - 2 n\mathcal{B} \right)}}{\left(1 - 2 n\mathcal{B}\right)^2}\right.\nn\\
&&\left.- \frac{e^{-i\left(1 + 2 n\mathcal{B} \right)}}{\left(1 + 2 n\mathcal{B}\right)^2}
\right]
-\mathcal{B}^{-1}\sum_{n=1}^{\infty}\left[
\frac{e^{-i\left(1 - 2 n\mathcal{B} \right)}}{1 - 2 n\mathcal{B}}\right.\nn\\
&&\left.- \frac{e^{-i\left(1 + 2 n\mathcal{B} \right)}}{1 + 2 n\mathcal{B}}
\right]\nn\\
&&=\frac{e^{i\left(2 - \mathcal{B}^{-1} \right)}}{\mathcal{B}\left(2\mathcal{B}-1\right)}{}_{2}F_{1}\left(1,1-\frac{1}{2\mathcal{B}};2-\frac{1}{2\mathcal{B}};e^{2i} \right)\nn\\
&&+ \frac{e^{i\left(-2 - \mathcal{B}^{-1} \right)}}{\mathcal{B}\left(2\mathcal{B}+1\right)}{}_{2}F_{1}\left(1,1+\frac{1}{2\mathcal{B}};2+\frac{1}{2\mathcal{B}};e^{-2i} \right)\nn\\
&&-\frac{i}{4\mathcal{B}^2}\left\{
e^{-i \left(\frac{1}{\mathcal{B}}+2\right)} \Phi \left(e^{-2 i},2,1+\frac{1}{2 \mathcal{B}}\right)\right.\nn\\
&&-\left.e^{i \left(-\frac{1}{\mathcal{B}}+2\right)} \Phi \left(e^{-2 i},2,1-\frac{1}{2 \mathcal{B}}\right)
\right\}
\end{eqnarray}
where ${}_{2}F_{1}(a,b;c;z)$ is the Hypergeometric function, while $\Phi(z,s,a)$ is
the Hurwitz-Lerch transcendent function, along with the exact infinite series expressions
\begin{eqnarray}
\sum_{n=1}^{\infty}\frac{e^{-i\left(1 + 2 n\mathcal{B}  \right)/\mathcal{B}}}{1 + 2 n\mathcal{B}} &=& \frac{e^{-i\left(2 + \mathcal{B}^{-1}  \right)}}{1 + 2\mathcal{B}}\\
&&\times{}_{2}F_{1}\left(1,1+\frac{1}{2\mathcal{B}};2+\frac{1}{2\mathcal{B}};e^{-2i} \right)\nn\\
\sum_{n=1}^{\infty}\frac{e^{-i\left(1 + 2 n\mathcal{B}  \right)/\mathcal{B}}}{\left(1 + 2 n\mathcal{B}\right)^2} &=&
\frac{e^{-i\left(2 + \mathcal{B}^{-1}  \right)}}{4\mathcal{B}^2}
\Phi\left(e^{-2i},2,
1+\frac{1}{2\mathcal{B}}\right)\nn
\end{eqnarray}

Combining these expressions, we obtain for the
mass shift contribution in this second subdominant kinematic region
\begin{eqnarray}
\left.\delta m_B^{(\pm)}\right|_{\mathcal{R}_2}
&&\sim \frac{2 m e^2}{(4\pi)^2}
\left\{
-1 - 2\mathcal{B}^{-1}\ln(\mathcal{B})
+ \mathcal{B}^{-1}\left(2\gamma\right.\right.\\
&&\left.\left.- 2\ln\left[|1 - e^{2i} | \right]
+ i\left(\pi \pm \frac{1}{2}\right)\right)
\right\}
+ O(\mathcal{B}^{-2}).\nn
\end{eqnarray}

\section{A periodic series expansion for
the functions $tan(x)$ and $cot(x)$}\label{AppD}
\begin{figure}
    \centering
    \includegraphics[scale=0.55]{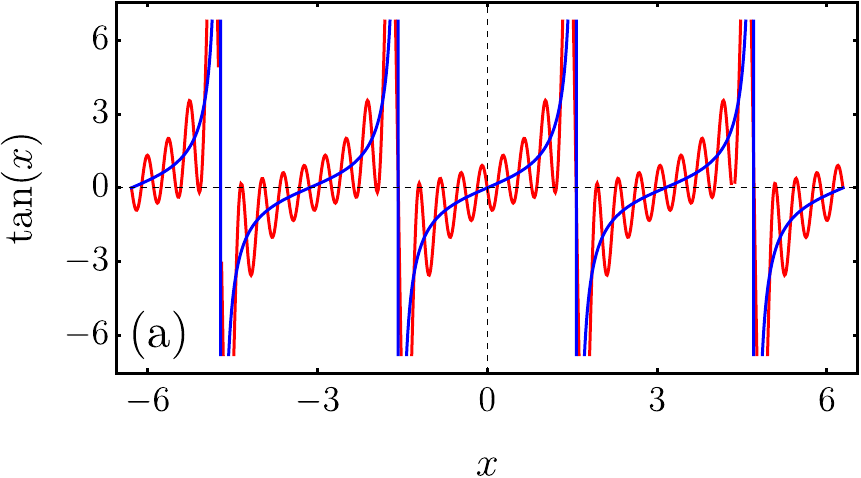}\nn\\
    \vspace{0.4cm}
    \includegraphics[scale=0.55]{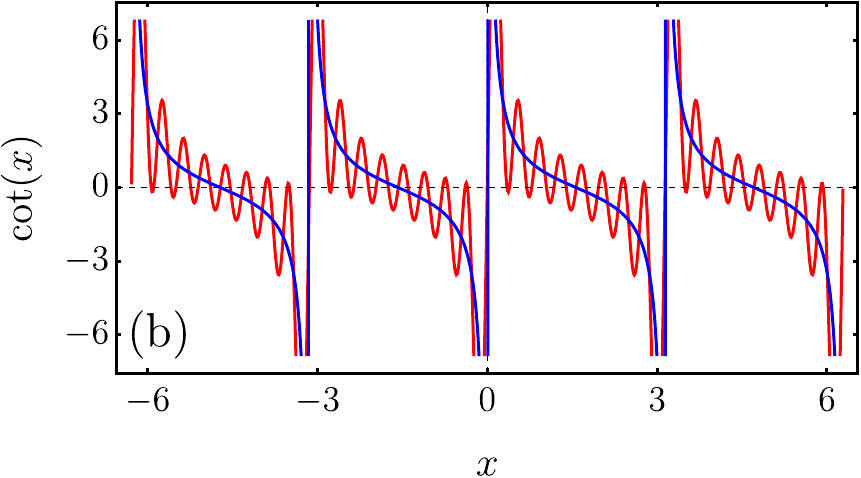}
    \caption{(a) Comparison between the periodic series Eq.~(\ref{eq:E4}) (truncated at 8 terms) and the function $\tan(x)$. (b) Comparison between the periodic series Eq.~(\ref{eq:E8}) and the function $\cot(x)$.}
    \label{Fig:tancotan}
\end{figure}

The function $\tan(z)$ possesses infinitely
many isolated poles over the real axis, at
every odd multiple of $\pi/2$, {\it i.e.} at
$z_j = (2j-1)\pi/2$, $j\in\mathbb{Z}$, and it is periodic with fundamental period $\pi$, $\tan(z + \pi) = \tan(z)$. Therefore, in the domain
of complex functions, it only admits a Laurent series representations defined inside concentric open discs
of the form $|z|<\pi/2$, $\pi/2<|z|<3\pi/2$, etc. It is therefore possible to obtain
an explicit representation within the open
real interval $\Re z = x \in (-\pi/2,\pi/2)$,
that extends periodically to all the 
open intervals of the form $(-(2j-1)\pi/2,(2j-1)\pi/2)$. For this purpose,
let us first consider the general definition in the complex plane
\begin{eqnarray}
\tan(z) &=& \frac{\sin(z)}{\cos(z)} = -i\frac{e^{iz}-e^{-iz}}{e^{iz} + e^{-iz}}\nonumber\\
&=& -\frac{d}{dz}\log\left( e^{iz} + e^{-iz} \right)
\label{eq:E1}
\end{eqnarray}
The complex $\log(z) = \ln|z| + i\arg(z)$ is an analytic function, and hence it possesses
a Taylor series with convergence radius $|z|\le 1$
\begin{eqnarray}
\log(1 + z) = \sum_{n=1}^{\infty}\frac{(-1)^{n-1}}{n}z^{n},\,\,\,|z|\le 1.
\label{eq:E2}
\end{eqnarray}
Therefore, using Eq.~(\ref{eq:E2}), we obtain the convergent series for $|e^{-2 i z}|\le 1$ (or equivalently for $\Im z \le 0$)
\begin{eqnarray}
\log\left( e^{iz} + e^{-iz} \right) &=&
\log\left(e^{iz}\left(1 + e^{-2iz} \right) \right)\nn\\
&=& i z + \log\left(1 + e^{-2iz} \right)\nn\\
&=& i z + \sum_{n=1}^{\infty}\frac{(-1)^{n-1}}{n}e^{-2 i n z}.
\label{eq:E3}
\end{eqnarray}
Choosing $z = x\in\mathbb{R}$ ($\Im z = 0$)
in the series~(\ref{eq:E3}), inserting into
Eq.~(\ref{eq:E1}), and futher taking the real part, we obtain the following periodic
series representation for the real tangent function
\begin{eqnarray}
\tan(x) &=& 2\sum_{n=1}^{\infty}(-1)^{n-1}\sin(2 n x)\nn\\
&=& i\sum_{n=1}^{\infty}(-1)^{n}\left( e^{2 i n x} - e^{-2 i n x} \right).
\label{eq:E4}
\end{eqnarray}
This constitutes a generalized Fourier series representation in the open interval $(-\pi/2,\pi/2)$, that also provides a periodic extension $\tan(x+\pi) = \tan(x)$. It must, however, be interpreted in the distributional sense, and not as a strict point wise convergence, since the last is only guaranteed
for the logarithm expansion in Eq.~(\ref{eq:E3}). Therefore, in the sense of distributions, for any continuous differentiable function $f(x)$ within an interval $x\in[a,b]$,
such that $a>-\pi/2$ and $b<\pi/2$,
we have
\begin{eqnarray}
\int_{a}^{b}f(x)\tan(x)dx
&=& -\Re \left[\int_{a}^{b}f(x)\frac{d}{dx}\log(e^{ix}+e^{-ix})dx\right]\nn\\
&=& \int_{a}^{b}\frac{df}{dx}(x)\Re\log(e^{ix} + e^{-ix})dx\nn\\
&-&\left[f(x) \log(e^{ix} + e^{-ix}) \right]_a^{b}
\label{eq:E5}
\end{eqnarray}
and hence the integral of the series provides the correct result (thanks to the point wise convergence of Eq.~(\ref{eq:E3})). This is illustrated in the upper panel of Fig.~ \ref{Fig:tancotan}.
Following a similar analysis, we have
that $\cot(z)$ is also a periodic function
with fundamental period $\pi$, {\it i.e.}
$\cot(z+\pi) = \cot(z)$, that possesses infinitely many poles located along the real
axis at $z_j = j \pi$, $j\in\mathbb{Z}$. Therefore, in the domain
of complex functions, it only admits a Laurent
series representations defined inside concentric
open discs of the form $0<|z|<\pi$, $\pi<|z|<2\pi$, etc. It is therefore possible
to obtain an explicit representation within the
open interval $\Re z = x \in (0,\pi)$, that
periodically extends to the remaining open
intervals of the form $(n\pi, (n+1)\pi)$. By
analogy with the previous case, we consider the definition in the complex plane of
\begin{eqnarray}
\cot(z) &=& \frac{\cos(z)}{\sin(z)} = i\frac{e^{iz}+e^{-iz}}{e^{iz} - e^{-iz}}\nn\\
&=& \frac{d}{dz}\log\left( e^{iz} - e^{-iz} \right)
\label{eq:E6}
\end{eqnarray}
Using again the convergent power series for the logarithm in Eq.~(\ref{eq:E3}), we have (for $\Im z \le 0$)
\begin{eqnarray}
\log\left( e^{iz} - e^{-iz} \right) &=&
i z + \sum_{n=1}^{\infty}\frac{(-1)^{n-1}}{n}
\left(- e^{-2iz}  \right)^{n}\nn\\
&=& i z - \sum_{n=1}^{\infty}\frac{1}{n}e^{-2inz}.
\label{eq:E7}
\end{eqnarray}
As before, setting $z = x\in \mathbb{R}$ ($\Im z = 0$), inserting into Eq.~(\ref{eq:E6}) and further taking the real part, we obtain the
periodic series representation for the
real $cot(x)$ function
\begin{eqnarray}
\cot(x) &=& 2\sum_{n=1}^{\infty}\sin(2 n x)\nn\\
&=& -i\sum_{n=1}^{\infty}\left( e^{2 i n x}
- e^{-2 i n x}
\right).
\label{eq:E8}
\end{eqnarray}
As in the former case, this series must be interpreted in the generalized sense of distributions, and not as point wise converging,
as follows from an identical analysis as in Eq.~(\ref{eq:E5}). This is illustrated in the lower panel of Fig.~\ref{Fig:tancotan}.

\section{Integrals of the incomplete Gamma functions}\label{AppE}
Here we show the details of the calculation of the integrals of the incomplete Gamma functions. For this we use the
exact series expansion
\begin{eqnarray}
\Gamma(0,iz) = -\gamma - \ln(iz) - \sum_{k=1}^{\infty}\frac{(-i z)^k}{k (k!)},
\label{eq:gamseries}
\end{eqnarray}
and integrate the three contributions separately.
Therefore, for the first integral we obtain
\begin{eqnarray}
&&\int_{\mathcal{B}^{-1}}^{1}\frac{dy}{y}(1 - y)(2 - y)\Gamma[0,i\mathcal{B}^{-1}(1-y)^2]
= 2\left[\ln(\mathcal{B})\right]^2\nn\\
&&+ \gamma\left\{\frac{5}{2}-3\mathcal{B}^{-1}
+\frac{1}{2}\mathcal{B}^{-2}-2 \ln (\mathcal{B})
\right\} - \frac{18 + i\pi}{4} + 5 \mathcal{B}^{-1}\nn\\
&&-\frac{1}{2}\mathcal{B}^{-2}
-2 \ln \left(i (1-\mathcal{B}^{-1})^2\right) \ln (\mathcal{B})\nn\\
&&+3 (1-\mathcal{B}^{-1} + \frac{1}{6}\mathcal{B}^{-2}) \ln \left(\frac{i (\mathcal{B}-1)^2}{\mathcal{B}^3}\right)\nn\\
&&+4 \text{Li}_2\left(1-\mathcal{B}^{-1}\right)-\ln (\mathcal{B}-1)+\frac{3}{2} \ln (\mathcal{B})\nn\\
&&- i\sum_{k=1}^{\infty}\frac{ \mathcal{B}^{-k} \left(-i(1-\mathcal{B}^{-1})\right)^{k+1}}{ (k+1) k k!}\nn\\
&&\times\left\{ \, _2F_1\left(1,2 (k+1);2 k+3;1-\mathcal{B}^{-1}\right)-\frac{1}{2}\right\}\nn\\
&=& 2\left[\ln(\mathcal{B})\right]^2 - \left( 2\gamma + \frac{5}{2} + i\pi\right)\ln(\mathcal{B}) + \left(3 + 2i \right)\mathcal{B}^{-1}\ln(\mathcal{B})\nn\\
&&+\frac{\left(30\gamma - 54 + 8\pi^2 + 15 i\pi\right)}{12}\nn\\
&&- \frac{\mathcal{B}^{-1}}{12}\left(48 + 36\gamma
+  i(18  \pi + 47)
\right) + O(\mathcal{B}^{-2}).
\end{eqnarray}
%For the second integral, after analogous procedure we have
%\begin{eqnarray}
%&&\int_{\mathcal{B}^{-1}}^{1}\frac{dy}{y}\Gamma[0,i\mathcal{B}^{-1}(1-y)^2] = \left[\ln(\mathcal{B})\right]^2
%-\gamma\ln(\mathcal{B})\\
%&&+ 2 \text{Li}_2\left(1-\mathcal{B}^{-1}\right)-\ln \left(i (1-\mathcal{B}^{-1})^2\right) \ln (\mathcal{B})\nn\\
%&&+\sum_{k=1}^{\infty}\frac{\left(-i \mathcal{B}\right)^{-k}}{k\,k!} \left[B_0(2 k+1,0)- B_{1-\mathcal{B}^{-1}}(2 k+1,0) \right]\nn\\
%&=& \left[\ln(\mathcal{B})\right]^2
%-\left(\gamma + \frac{i\pi}{2}\right)\ln(\mathcal{B})
%+ \frac{\pi^2}{3}+i\mathcal{B}^{-1}\ln(\mathcal{B})\nn\\
%&&-\mathcal{B}^{-1}\frac{\left(4 + 3 i\right)}{2} + O(\mathcal{B}^{-2}),
%\end{eqnarray}
%where $B_z(a,b)$ is the Incomplete Beta function.

Finally, for the third type of integrals, with $n = 1,\ldots,\infty$, we obtain from Eq.~(\ref{eq:gamseries})
\begin{eqnarray}
&&\int_{\mathcal{B}^{-1}}^{1}dy\Gamma[0,i\mathcal{B}^{-1}(1-y)(1 -y + 2 n \mathcal{B})]
= -\gamma\ln(\mathcal{B})\nn\\
&&-\int_{\mathcal{B}^{-1}}^{1}dy \ln\left[i\mathcal{B}^{-1}(1-y)(1 -y + 2 n \mathcal{B})  \right]
- S(\mathcal{B},n).\nn\\
\label{eq:SB1}
\end{eqnarray}

Here, we defined the series
\begin{eqnarray}
S(\mathcal{B},n) &=&
\sum_{k=1}^{\infty}\frac{(i\mathcal{B})^{-k}}{k\,k!}\int_{\mathcal{B}^{-1}}^{1}
dy\left[(1-y)(1-y+2 n\mathcal{B}) \right]^{k}\nn\\
\label{eq:SBN}
\end{eqnarray}
Each of the integrals can be performed analytically, to obtain
\begin{eqnarray}
\int_{\mathcal{B}^{-1}}^{1}
&&dy\left[(1-y)(1-y+2 n\mathcal{B}) \right]^{k}
= \frac{2^k (\mathcal{B}-1)^{k+1} n^k }{\mathcal{B}(k + 1)}\nn\\
&&\times \, _2F_1\left(-k,k+1;k+2;\frac{1-\mathcal{B}}{2 \mathcal{B}^2 n}\right)
\end{eqnarray}

Here, 
$_2F_1\left(a_1,a_2;b_1;z\right)$ is one of the Hypergeometric functions. Inserting this result
into Eq.~(\ref{eq:SB1}), we obtain
\begin{eqnarray}
S(\mathcal{B},n) &=& \left(1 - \mathcal{B}^{-1} \right) \sum_{k=1}^{\infty}\frac{\left(-2 n i  \left(1 - \mathcal{B}^{-1} \right) \right)^k}{k (k+1)!}\nn\\
&&\times\, _2F_1\left(-k,k+1;k+2;\frac{1-\mathcal{B}}{2 \mathcal{B}^2 n}\right).
\label{eq:SBN2}
\end{eqnarray}
Since the Hypergeometric function  satisfies
\begin{eqnarray}
\!\!\!\!\!\!\! _2F_1\left(-k,k+1;k+2;\frac{1-\mathcal{B}}{2 \mathcal{B}^2 n}\right) = 1
+ O(\mathcal{B}^{-1}),
\end{eqnarray}
and when substituted into Eq.~(\ref{eq:SBN2}), we thus
obtain the asymptotics
\begin{eqnarray}
S(\mathcal{B},n) &=& \left(1 - \mathcal{B}^{-1} \right) \sum_{k=1}^{\infty}\frac{\left(-2 n i  \left(1 - \mathcal{B}^{-1} \right) \right)^k}{k (k+1)!} + O(\mathcal{B}^{-2})\nn\\
&=& -\frac{e^{-2 i n}}{{2 n}} \left(-i e^{2 i n}-2 e^{2 i n} n+2 \gamma  e^{2 i n} n \right.\nn\\
&&\left.+2 e^{2 i n} n \ln (2 i n)+2 e^{2 i n} n \Gamma (0,2 i n)+i\right)\nn\\
&+&\mathcal{B}^{-1}\left(\ln (2 i n)+\Gamma (0,2 i n)+\gamma\right)+ O(\mathcal{B}^{-2})
\end{eqnarray}

Using this result into Eq.~(\ref{eq:SB1}), as shown in the main text we need the combination
\begin{eqnarray}
&&\int_{\mathcal{B}^{-1}}^{1}dy\left(
\Gamma[0,i\mathcal{B}^{-1}(1-y)(1 -y - 2 n \mathcal{B})]\right.\nn\\
&&\left.- \Gamma[0,i\mathcal{B}^{-1}(1-y)(1 -y + 2 n \mathcal{B})]\right) = \nn\\
&&-\int_{\mathcal{B}^{-1}}^{1}dy
\left( \ln\left[i\mathcal{B}^{-1}(1-y)(1 -y - 2 n \mathcal{B})  \right]\right.\nn\\
&&\left.- \ln\left[i\mathcal{B}^{-1}(1-y)(1 -y + 2 n \mathcal{B})  \right] \right)+ S(\mathcal{B},n) - S\left(\mathcal{B},-n \right)\nn\\
&&= \left(1 - \mathcal{B}^{-1}  \right)\left( \Gamma[0,2in] - \Gamma[0,-2in] \right)\nn\\
&&- \frac{i}{n}
+\frac{i}{2 n}\left( e^{2 i n} - e^{-2 i n} \right)+ O(\mathcal{B}^{-2}).
\end{eqnarray}
Finally, using the Taylor expansion for the logarithm
\begin{eqnarray}
\ln(1 + z) = \sum_{n=1}^{\infty}\frac{(-1)^{n+1}z^n}{n},
\end{eqnarray}
we have the simple identities
\begin{subequations}
\begin{eqnarray}
\sum_{n=1}^{\infty}\frac{(-1)^n}{n} = -\ln(2)
\eea
\bea
&&\sum_{n=1}^{\infty}\frac{(-1)^n}{n}\left( e^{2 i n} - e^{-2 i n}  \right)\nn\\
&=&
-\ln\left( 1 + e^{2i} \right)+ \ln\left( 1 + e^{-2i} \right)\nn\\
&=& -\ln\left( 1 + e^{2i} \right) + \ln\left(e^{-2i}\left( 1 + e^{2i}\right) \right)\nn\\
&=& - 2 i,
\end{eqnarray}
\end{subequations}

along with the sum
\begin{eqnarray}
\sum_{n=1}^{\infty}(-1)^n\left(\Gamma[0,2in] - \Gamma[0,-2in] \right) = 0.421794 i.
\end{eqnarray}
Therefore, we finally obtain
\begin{eqnarray}
&&\pm i\sum_{n=1}^{\infty}\int_{\mathcal{B}^{-1}}^{1}dy\left\{\Gamma[0,i\mathcal{B}^{-1}(1-y)(1 -y - 2 n \mathcal{B})]\right.\\
&&\left.-
\Gamma[0,i\mathcal{B}^{-1}(1-y)(1 -y + 2 n \mathcal{B})]\right\}
\nn\\
&&= \mp 0.421794\left(1-\mathcal{B}^{-1} \right) \mp \ln(2) \pm i\left(1-\mathcal{B}^{-1}\right)\frac{\ln(2)}{2}\nn\\
&&+ O(\mathcal{B}^{-2}).\nn
\label{eq:sumgamma}
\end{eqnarray}

\section{Connection with Ref.~\cite{Tsai:1974df}}\label{Ap_Tsaisconnection}
In this appendix we compare our results with well-known calculations in the limit of a strong magnetic field. In particular, we demonstrate that the expressions for the counter-terms and the relevant integrals are the same as the ones presented in Ref.~\cite{Tsai:1974df} by Tsai. In order to do that, let us begin with Eqs.~(\ref{eq:ct1}) and~(\ref{eq:ct2}) which correspond to the counter-terms, namely
\bea
c.t._1=-(2-y),
\label{ct1}
\eea
and
\bea
c.t._2=-(\slashed{\rho}-1)\left[-\frac{y}{m}+2i\frac{sy(1-y)(2-y)}{m}\right],
\label{ct2}
\eea
so that, in terms of the canonical momentum $p^\mu$, the sum of counter-terms is
\bea
c.t._1+c.t._2&=&-(2-y)\nn\\
&+&\frac{\slashed{p}-m}{m}\left[\frac{y}{m}-2i\frac{sy(1-y)(2-y)}{m}\right].\nn\\
\label{c1c2}
\eea

In order to compare with Ref.~\cite{Tsai:1974df}, it is necessary to identify our integration variables with the ones used in that work. Explicitly, the latter is given by
\bea
u&\to& 1-y\nn\\
s&\to&m^2s.
\label{identifications1}
\eea
Moreover, in Ref.~\cite{Tsai:1974df} the metric is
\bea
g^{\mu\nu}=\text{diag}(-1,1,1,1)
\eea
whereas we use
\bea
g^{\mu\nu}=\text{diag}(1,-1,-1,-1),
\eea 
therefore, we also identify
\bea
\slashed{p}+m\to\slashed{p}-m.
\label{identifications2}
\eea
Now, the counter-term in Ref.~\cite{Tsai:1974df} is
\bea
c.t.=-(1-u)-(m+\slashed{\Pi})\left[\frac{1-u}{m}-2imu(1-u^2)s\right],\nn\\
\eea
so that after applying Eqs.~(\ref{identifications1}) and~(\ref{identifications2}) it gives Eq.~(\ref{c1c2}), with the difference that in our calculation there appears $\slashed{p}$ instead of $\slashed{\Pi}$.

For the sake of completeness, let us find the parallelism with the other expressions in the reference of interest. First, for Eq.~(21) in Tsai's work:
\bea
\tilde{M}(p)&=&\frac{\alpha}{2\pi}\int\frac{ds}{s}du\frac{e^{-is(um^2+\varphi)}e^{i\zeta \y}}{(1-u)\cos\y+u(\sin \y)/\y}\nn\\
&\times&\left[1+e^{-2i\zeta \y}+(1-u)e^{-2i\zeta \y}\frac{\slashed{p}_\parallel}{m}\right.\nn\\
&+&\left.\frac{(1-u)\slashed{p}_\perp/m}{(1-u)\cos\y+u(\sin\y)/\y}\right],
\label{MTsai}
\eea
where $\y$ means the $y$-variable used there ($\y=eHsu$) and $\zeta=q\hat{\Sigma}_3=q\gamma^1\gamma^2$ is the spin matrix. Now, from the fact that
\bea
e^{i\zeta\y}&=&\cos\y+iq\hat{\Sigma}_3\sin\y\nn\\
&\to& e^{i\zeta\y}+e^{-i\zeta\y}=2\cos\y,
\eea
equation~(\ref{MTsai}) becomes:
\bea
\tilde{M}(p)&=&\frac{\alpha}{2\pi}\int\frac{ds}{s}du\frac{e^{-is(um^2+\varphi)}e^{i\zeta \y}}{(1-u)\cos\y+u(\sin \y)/\y}\nn\\
&\times&\left[2\cos\y+(1-u)\left(\cos\y-i\zeta\sin\y\right)\frac{\slashed{p}_\parallel}{m}\right.\nn\\
&+&\left.\frac{(1-u)\slashed{p}_\perp/m}{(1-u)\cos\y+u(\sin\y)/\y}\right].
\label{M2}
\eea

On the other hand, from Eq.~(20) in Tsai's work
\bea
\varphi=u(1-u)p^2_\parallel+\frac{u}{\y}\frac{(1-u)\sin\y}{(1-u)\cos\y+u(\sin\y)/\y}p^2_\perp,\nn\\
\eea
then, the overall phase factor is
\bea
-is(um^2+\varphi)&=&-ism^2\left[u+u(1-u)\frac{p^2_\parallel}{m^2}\right.\\
&+&\left.\frac{u}{\y}\frac{(1-u)\sin\y}{(1-u)\cos\y+u(\sin\y)/\y}\frac{p^2_\perp}{m^2}\right],\nn
\label{overallphasefactor}
\eea
but given the metric choice, $p^2_\parallel$ connects with our $\rho^2_\parallel$ as
\bea
p^2_\parallel=-m^2\rho^2_\parallel.
\eea

Moreover, by using Eqs.~(\ref{identifications1}) and~(\ref{identifications2}) 
\bea
\y&=&eHsu\to\mathcal{B}s(1-y),\nn\\
\frac{u}{\y}\sin\y&\to&\frac{\sin\left[\mathcal{B}s(1-y)\right]}{\mathcal{B}s},\nn\\
(1-u)\cos\y&\to&y\cos\left[\mathcal{B}s(1-y)\right].
\eea

The above replacements in Eq.~(\ref{overallphasefactor}) yield our phase factor. With the same argument, note that the factors $(A),\,(B)$ and $(C)$ given in our Eqs.~(\ref{TermA})-(\ref{TermC}), can be identified in Eq.~(\ref{M2}) as
\bea
(A)&\to&\frac{2\cos\y+(1-u)\cos\y}{(1-u)\cos\y+u(\sin\y)/\y}\frac{\slashed{p}_\parallel}{m}\nn\\
(B)&\to&\frac{(1-u)}{\left[(1-u)\cos\y+u(\sin\y)/\y\right]^2}\frac{\slashed{p}_\parallel}{m}\nn\\
(C)&\to&-i\zeta\frac{(1-u)\sin\y}{(1-u)\cos\y+u(\sin\y)/\y}\frac{\slashed{p}_\perp}{m},
\eea
with $\zeta=q\hat{\Sigma}_3=q\gamma^1\gamma^2$.

% \bibliographystyle{spphys}       % APS-like style for physics
% %\bibliography{}   % name your BibTeX data base

% \bibliography{bibliografia}% Produces the bibliography via BibTeX.
\bibliography{BibQEDVertex.bib}
% \begin{thebibliography}{89}
% %On gauge invariance and vacuum polarization
% \bibitem{Schwinger:1951nm} J. S. Schwinger, Phys. Rev. {\bf 82}, 664--679 (1951).

% %Dimensional reduction and catalysis of dynamical symmetry breaking by a magnetic field
% \bibitem{Gusynin:1995nb} V. P. Gusynin, V. A. Miransky and I. A. Shovkovy, Nucl. Phys. B {\bf 462}, 249--290 (1996)

% %Schwinger-Dyson equation approach to chiral symmetry breaking in an external magnetic field
% \bibitem{Leung:1995mh} C. N. Leung, Y.J. Ng and A.W. Ackley, Phys. Rev. D {\bf 54}, 4181--4184 (1996).

% %Dynamical chiral symmetry breaking by a magnetic field in QED
% \bibitem{Gusynin:1995gt} V.P. Gusynin, V.A. Miransky, I.A. Shovkovy, Phys. Rev. D {\bf 52}, 4747--4751 (1995).

% \end{thebibliography}

\end{document}